\documentclass[longauth]{aa}  
\usepackage{ulem}
\usepackage{graphicx}
\usepackage{txfonts}
\usepackage[colorlinks=true,linkcolor=red,anchorcolor=blue,citecolor=blue,filecolor=black,menucolor=black,runcolor=black,urlcolor=blue]{hyperref}

\newcommand{\arcs}{$\arcsec$\xspace}
\newcommand{\CII}{[\ion{C}{II}]158$\mu$m}
\newcommand{\oddpm}[2]{\raisebox{0.5ex}{\tiny$\substack{+#1 \\ -#2}$}}

\begin{document}


 
    \title{MIRI/JWST observations reveal an extremely obscured starburst in the z=6.9 system SPT0311-58} 
   

   \author{J. \'Alvarez-M\'arquez\inst{\ref{inst:CAB}} \and A. Crespo G\'omez\inst{\ref{inst:CAB}} \and L. Colina\inst{\ref{inst:CAB}} \and M. Neeleman\inst{\ref{inst:MPIA}} \and F. Walter\inst{\ref{inst:MPIA}} \and A. Labiano\inst{\ref{inst:CAB},\ref{inst:Telespazio}} \and P. P\'erez-Gonz\'alez\inst{\ref{inst:CAB}} \and A. Bik\inst{\ref{inst:Stockholm}}  \and H.U. Noorgaard-Nielsen\inst{\ref{inst:DTU}}\thanks{Deceased} \and G. Ostlin\inst{\ref{inst:Stockholm}} \and G. Wright\inst{\ref{inst:UKATC}} \and A. Alonso-Herrero\inst{\ref{inst:CAB-ESAC}} \and R. Azollini\inst{\ref{inst:CAB}, \ref{inst:Dublin}} \and K.I. Caputi\inst{\ref{inst:Groningen},\ref{inst:DAWN}} \and A. Eckart\inst{\ref{inst:Köln}} \and O. Le F\`evre\inst{\ref{inst:LAM}}$^{\star}$ \and M. Garc\'ia-Mar\'in\inst{\ref{inst:ESA}} \and T.R. Greve\inst{\ref{inst:DTU}, \ref{inst:UCL}} \and J. Hjorth\inst{\ref{inst:DARK}} \and O. Ilbert\inst{\ref{inst:LAM}} \and S. Kendrew\inst{\ref{inst:ESA}}  \and J.P. Pye\inst{\ref{inst:Leicester}} \and T. Tikkanen\inst{\ref{inst:Leicester}} 
   \and M. Topinka\inst{\ref{inst:Dublin}} 
   \and P. van der Werf\inst{\ref{inst:Leiden}} \and M. Ward\inst{\ref{inst:Durham}}
   \and E. F. van Dishoeck\inst{\ref{inst:Leiden}} 
   \and M. G\"udel\inst{\ref{inst:Vienna}, \ref{inst:MPIA}, \ref{inst:ETH}} \and Th. Henning\inst{\ref{inst:MPIA}} \and P.O. Lagage\inst{\ref{inst:AIM}} \and T. Ray\inst{\ref{inst:Dublin}} \and C. Waelkens\inst{\ref{inst:Leuven}}}

   \institute{Centro de Astrobiolog\'{\i}a (CAB), CSIC-INTA, Ctra. de Ajalvir km 4, Torrej\'on de Ardoz, E-28850, Madrid, Spain\\  \email{jalvarez@cab.inta-csic.es} \label{inst:CAB}
    \and Max-Planck-Institut f\"ur Astronomie, K\"onigstuhl 17, 69117 Heidelberg, Germany\label{inst:MPIA}
    \and Telespazio UK for the European Space Agency, ESAC, Camino Bajo del Castillo s/n, 28692 Villanueva de la Ca\~{n}ada, Spain \label{inst:Telespazio}
    \and Department of Astronomy, Stockholm University, Oscar Klein Centre, AlbaNova University Centre, 106 91 Stockholm, Sweden \label{inst:Stockholm}
    \and Cosmic Dawn Center, DTU Space, Technical University of Denmark, Elektrovej 327, 2800 Kgs. Lyngby, Denmark \label{inst:DTU}
    \and UK Astronomy Technology Centre, Royal Observatory Edinburgh, Blackford Hill, Edinburgh EH9 3HJ, UK \label{inst:UKATC} 
    \and Centro de Astrobiolog\'ia (CAB), CSIC-INTA, Camino Viejo del Castillo s/n, 28692 Villanueva de la Ca\~{n}ada, Madrid, Spain \label{inst:CAB-ESAC}
    \and Dublin Institute for Advanced Studies, Astronomy \& Astrophysics Section, 31 Fitzwilliam Place, Dublin 2, Ireland \label{inst:Dublin}
    \and Kapteyn Astronomical Institute, University of Groningen, P.O. Box 800, 9700 AV Groningen, The Netherlands \label{inst:Groningen} 
    \and Cosmic Dawn Centre, Copenhagen, Denmark \label{inst:DAWN}
    \and I.Physikalisches Institut der Universit\"at zu K\"oln, Z\"ulpicher Str. 77, 50937 K\"oln, Germany \label{inst:Köln}
    \and Aix Marseille Universit\'e, CNRS, LAM (Laboratoire d’Astrophysique de Marseille) UMR 7326, 13388, Marseille, France \label{inst:LAM}
    \and European Space Agency, Space Telescope Science Institute, Baltimore, Maryland, USA \label{inst:ESA}
    \and Department of Physics and Astronomy, University College London, Gower Place, London WC1E 6BT, UK \label{inst:UCL}
    \and DARK, Niels Bohr Institute, University of Copenhagen, Jagtvej 128, 2200 Copenhagen, Denmark \label{inst:DARK}
    \and School of Physics \& Astronomy, Space Research Centre, Space Park Leicester, University of Leicester, 92 Corporation Road, Leicester, LE4 5SP, UK \label{inst:Leicester} 
    \and Leiden Observatory, Leiden University, PO Box 9513, 2300 RA Leiden, The Netherlands \label{inst:Leiden}
    \and Centre for Extragalactic Astronomy, Durham University, South Road, Durham DH1 3LE, UK \label{inst:Durham}
    \and University of Vienna, Department of Astrophysics, Türkenschanzstrasse 17, 1180 Vienna, Austria \label{inst:Vienna}
    \and AIM, CEA, CNRS, Universit\`e Paris-Saclay, Universit\`e Paris Diderot, Sorbonne Paris Cit\`e, F-91191 Gif-sur-Yvette, France \label{inst:AIM}
    \and Institute of Astronomy, KU Leuven, Celestijnenlaan 200D bus 2401, 3001 Leuven, Belgium \label{inst:Leuven}
    \and Institute of Particle Physics and Astrophysics, ETH Zurich, Wolfgang-Pauli-Str 27, 8093 Zurich, Switzerland \label{inst:ETH} 
    }

   \date{Received ; accepted}


 

\abstract
{
Luminous infrared starbursts in the early Universe are thought to be the progenitors of massive quiescent galaxies identified at redshifts 2 to 4. Using the Mid-IRfrared Instrument (MIRI) on board the \textit{James Webb Space Telescope} (JWST), we present mid-infrared sub-arcsec imaging and spectroscopy of such a starburst: the slightly lensed hyper-luminous infrared system SPT0311-58 at $z$=6.9. The MIRI IMager (MIRIM) and Medium Resolution Spectrometer (MRS) observations target the stellar (rest-frame 1.26$\mu$m emission) structure and ionised (Pa$\alpha$ and H$\alpha$) medium on kpc scales in the system. The MIRI observations are compared with existing ALMA far-infrared continuum and [\ion{C}{II}]158$\mu$m imaging at a similar angular resolution. Even though the ALMA observations imply very high star formation rates (SFRs) in the eastern (E) and western (W) galaxies of the system, the H$\alpha$ line is, strikingly, not detected in our MRS observations. This fact, together with the detection of the ionised gas phase in Pa$\alpha$, implies very high internal nebular extinction with lower limits ($A_\mathrm{V}$) of 4.2 (E) and 3.9 mag (W) as well as even larger values (5.6 (E) and 10.0 (W)) by spectral energy distribution (SED) fitting analysis. The extinction-corrected Pa$\alpha$ lower limits of the SFRs are 383 and 230\,M$_\mathrm{\odot}$\,yr$^{-1}$ for the E and W galaxies, respectively. This represents 50\% of the SFRs derived from the [\ion{C}{II}]158$\mu$m line and infrared light for the E galaxy and as low as 6\% for the W galaxy. The MIRIM observations reveal a clumpy stellar structure, with each clump having  3 to 5 $\times$10$^{9}$\,M$_\mathrm{\odot}$ mass in stars, leading to a total stellar mass of 2.0 and 1.5$\times$10$^{10}$\,M$_\mathrm{\odot}$ for the E and W galaxies, respectively. The specific star formation (sSFR) in the stellar clumps ranges from 25 to 59\,Gyr$^{-1},$ assuming a star formation with a 50-100 Myr constant rate. This sSFR is three to ten times larger than the values measured in galaxies of similar stellar mass at redshifts 6 to 8. Thus, SPT0311-58 clearly stands out as a starburst system when compared with typical massive star-forming galaxies at similar high redshifts. The overall gas mass fraction is $M_\mathrm{gas}$/$M_*\sim\,3$, similar to that of $z${}$\sim$4.5-6 star-forming galaxies, suggesting a flattening of the gas mass fraction in massive starbursts up to redshift 7. The kinematics of the ionised gas in the E galaxy agrees with the known [\ion{C}{II}] gas kinematics, indicating a physical association between the ionised gas and the cold ionised or neutral gas clumps. The situation in the W galaxy is more complex, as it appears to be a velocity offset by about +700\,km\,s$^{-1}$ in the Pa$\alpha$ relative to the [\ion{C}{II}] emitting gas. The nature of this offset and its reality are not fully established and require further investigation. The observed properties of SPT0311-58, such as the clumpy distribution at sub(kpc) scales and the very high average extinction, are similar to those observed in low- and intermediate-$z$ luminous (E galaxy) and ultra-luminous (W galaxy) infrared galaxies, even though SPT0311-58 is observed only $\sim$800\,Myr after the Big Bang. Such massive, heavily obscured clumpy starburst systems as SPT0311-58 likely represent the early phases in the formation of a massive high-redshift bulge, spheroids and/or luminous quasars. This study demonstrates that MIRI and JWST are, for the first time, able to explore the rest-frame near-infrared stellar and ionised gas structure of these galaxies, even during the Epoch of Reionization.}

\titlerunning{MIRI/JWST study of SPT0311-58}
\maketitle

\section{Introduction}
\label{sec:1.}
        
Since the discovery of massive quiescent galaxies at $z$=2$-$4 \citep{Cimatti+04,Cimatti+06,vanDokkum+04,Stockmann+20,Valentino+20}, it has been hypothesised that their progenitors in the early Universe must be short-lived massive starbursts at $z$>4 \citep{Casey+14,Toft+14}. Analysis of Atacama Large Millimeter Array (ALMA) observations of far-infrared   ultra-red sources has identified galaxies at $z$=3$-$7 \citep{Iono+16,Oteo+16,Oteo+17,Gomez-Guijarro+18,Marrone+18} with extreme infrared luminosities ($\log{L_\mathrm{IR}}$(L$_\mathrm{\odot}$)\,$>$\,13) and star formation rates (SFRs; SFR$>$10$^3$\,M$_\mathrm{\odot}$\,yr$^{-1}$) in very compact regions (1$-$2\,kpc radius). These galaxies are the most luminous and intense star-forming galaxies in the Universe. They are even identified as maximal starbursts (i.e. $\Sigma_\mathrm{SFR}$ $\approx$ 0.5$-$1x10$^3$ \,M$_\mathrm{\odot}$\,yr$^{-1}$\,kpc$^{-2}$ and $\Sigma_{L_\mathrm{IR}}$ $\approx$ 10$^{13}$ \,L$_\mathrm{\odot}$\,kpc$^{-2}$; \citealt{Thompson+05,Crocker+18}). The most accepted evolutionary scenario is that before the galaxy became a naked quasi-stellar object and finally a compact quiescent galaxy  \citep{Hopkins+08a,Hopkins+08b}, interactions and/or mergers of gas- and dust-rich galaxies in the early Universe ($z$>4) went through an initial short (<100\,Myr) and intense (SFR$>$10$^3$\,M$_\mathrm{\odot}$\,yr$^{-1}$) dusty starburst phase at coalescence, resulting in the expulsion (blowout phase) of the remaining interstellar medium (ISM). This basic evolutionary  scenario has recently been challenged, as the comoving density of detected infrared-luminous dusty star-forming galaxies at $z$>4  appears to be lower than current estimates of quiescent galaxies at $z${}$\sim$3$-$4 \citep{Valentino+20}. In addition, recent ALMA-based studies indicate that dusty star-forming galaxies at $z$>4 have a wide range of underlying morphologies and physical origins: they can be members of protoclusters \citep{Oteo+17,Drake+20}, they can be involved in major and/or minor mergers \citep{Riechers+20,Ginolfi+20,Gomez-Guijarro+18}, or they can be large star-forming discs \citep{Hodge+12,Jimenez-Andrade+20}. Therefore, the triggering mechanisms of these extremely massive and luminous starbursts and their subsequent evolution is far from clear. Shedding new light on this topic requires use of the \textit{James Webb Space Telescope's} (JWST) high angular resolution imaging of their stellar structure and close environment.

So far, most of the knowledge about the structure of dusty star-forming galaxies at $z$>4 comes from the ALMA dust continuum and [\ion{C}{II}]158$\mu$m emission line maps. These studies conclude that while the dust emission is compact, the [\ion{C}{II}]158$\mu$m emission tracing the bulk of the neutral phases (atomic and/or molecular) is more extended, with an average effective radius of 1 and 1.7\,kpc, respectively \citep{Cooke+18}. With an SFR of $\sim$1000\,M$_\mathrm{\odot}$\,yr$^{-1}$ in such small volumes, a compact massive core would form in a very short time \citep[$<$100\,Myr,][]{Toft+14}, as seen in intermediate redshift ($z${}$\approx$2) quiescent galaxies. Due to their extreme faintness at (rest-frame) optical wavelengths, the stellar structure of extreme starbursts is mostly unknown and out of reach with \textit{Hubble Space Telescope} (HST) and 8-10 m telescopes. Only a handful of galaxies have IRAC/Spitzer detections in the mid-infrared,  but these lack the angular resolution to investigate their structure. The JWST, with its combination of subarcsec angular resolution and sensitivity orders of magnitudes better than any other previous telescope in the near- and mid-infrared spectral range (i.e. 1 to 28 $\mu$m), opens for the first time the possibility of investigating the structure of the stellar light and ionised gas in these galaxies on (sub)kpc scales. 

SPT0311-58 is the highest redshift ($z$=6.9) infrared-luminous dusty star-forming system known so far \citep{Strandet+17,Marrone+18} and one of the only two identified in the Epoch of Reionization (the other is HFLS3 at a redshift of 6.3; \citealt{Riechers+13}). This system was first discovered in the South Pole Telescope Survey \citep{Vieira+13} and was later studied in detail with ALMA, tracing several molecular and atomic lines at different angular resolutions from 0.3$-$0.5 to 0.07$-$0.08 arcsec \citep{Marrone+18,Jarugula+21,Spilker+22}. The system consists of an East-lensed (E) and a West-lensed (W) galaxy that are separated by a projected distance of 8\,kpc, relative velocities of 700 km s$^{-1}$ and average lensing magnification of 1.3 (E) and 2.2 (W). Their intrinsic infrared luminosity (8 to 1000\,$\mu$m) places the two galaxies of the system in the range of ultra-luminous (E) and hyper-luminous (W) infrared galaxies, with luminosities of 3.5$\pm$0.7$\times$10$^{12}$\,L$_\mathrm{\odot}$ and 26$\pm$12$\times$10$^{12}$\,L$_\mathrm{\odot}$, respectively \citep{Jarugula+21}. SPT0311-58 is a very gas-rich system with intrinsic total gas masses (E and W) of 3.1$\pm$2.7$\times$10$^{10}$\,M$_\mathrm{\odot}$ and 5.4$\pm$3.4$\times$10$^{11}$\,M$_\mathrm{\odot}$ \citep{Jarugula+21}. Star formation is proceeding at a very rapid pace, with  [\ion{C}{II}]158$\mu$m-based rates of 778 and 1775\,M$_\mathrm{\odot}$\,yr$^{-1}$ in the E and W galaxies, respectively \citep{Spilker+22}. This star formation appears to be distributed across several clumps with sizes of about 1\,kpc radius, and the stars form at rates of 100 to 400\,M$_\mathrm{\odot}$\,yr$^{-1}$. The kinematics of the [\ion{C}{II}] emitting gas indicate that  neither of the two galaxies in SPT0311-58 agree with the velocity pattern from a thin rotation-supported disc. The perturbed kinematics are rather likely produced by the combined effect of tidal interactions, disc gravitational fragmentation, and increased turbulence  associated with the star formation process itself \citep{Spilker+22}.

This paper presents mid-infrared imaging and spectroscopy obtained with the JWST Mid-Infrared Instrument (MIRI) \citep{Rieke+15, Wright+15}. For the first time, these data show  the stellar structure and ionised gas on kpc scales of a massive star-forming system from when the Universe was only 800\,Myr old. The structure of the paper is as follows. Section~\ref{sec:2.} presents the JWST data, including the calibration and post-calibration processing ($\S$\ref{sec:2.1.} to $\S$\ref{sec:2.3.}). The ALMA and HST ancillary data are presented together with details about the relative astrometry of the different sets of data ($\S$\ref{sec:2.4.} and $\S$\ref{sec:2.5.}). Section~\ref{sec:3.} presents the removal of the lensing galaxy ($\S$\ref{sec:3.1.}), the MIRI photometry ($\S$\ref{sec:3.2.}), and spectroscopic ($\S$\ref{sec:3.3.}) measurements. Section~\ref{sec:4.} contains our results and discussion, including the overall spectral energy distribution (SED) of SPT0311-58 components ($\S$\ref{sec:4.1.}), the stellar structure of the system ($\S$\ref{sec:4.2.}), the internal extinction based on the hydrogen recombination lines and SED fit ($\S$\ref{sec:4.3.}), the kinematics of the ionised medium ($\S$\ref{sec:4.4.}), the starburst nature and SFRs ($\S$\ref{sec:4.5.}), the ratio of the stellar to dynamical mass in this system ($\S$\ref{sec:4.6.}), and the SPT0311-58 system in the context of cosmological models ($\S$\ref{sec:4.7.}).

SPT0311-58 is a slightly magnified lensed system. Throughout the paper, average lensing factors of 1.3 and 2.2 for the E and W galaxies, respectively, of the SPT0311-58 system \citep{Marrone+18,Jarugula+21} are used when converting observed fluxes to intrinsic fluxes as well as luminosities, and derived properties, such as SFRs and stellar masses. We assume a $\Lambda$-Cold Dark Matter ($\Lambda$CDM) cosmology model with $\Omega_{m}$=0.310 and H$_0$=67.7\,km\,s$^{-1}$\,Mpc$^{-1}$ \citep{PlanckCollaboration18VI}. For this cosmology, 1 arcsec corresponds to 5.39\,kpc at $z$=6.9, and the luminosity distance is $D_\mathrm{L}$=69\,Gpc. Following \citet{Madau&Dickinson+14}, the SFRs and stellar masses from published data based on Salpeter initial mass function (IMF) \citep{Salpeter+55} were transformed into Kroupa \citep{Kroupa+01} and Chabrier \citep{Chabrier+03} by multiplying by the same factors, 0.67 (SFR) and 0.66 ($M_\mathrm{*}$).

\section{Observations, calibration, and data processing}
\label{sec:2.}

\subsection{JWST data}
\label{sec:2.1.}

SPT0311-58 JWST data were obtained on July 17, 2022, using the MIRI as part of the European Consortium MIRI Guaranteed Time (proposal ID 1264). The data is composed of imaging observations from the MIRI Imager \citep[MIRIM;][]{Bouchet+15} and integral field spectroscopic observations from the medium resolution spectrograph \citep[MRS;][]{Wells+15,Labiano+21}. The image was taken with the F1000W filter centred at 10$\mu$m for a total integration time of 2719.5 seconds in FASTR1 read-out mode and using a four-point dither pattern optimised for point source. Each dither was divided into three integrations of 81 groups. The MRS data were taken in SLOWR1 read-out mode with two independent configurations, SHORT and MEDIUM, and each had a total integration time of 7549.2 seconds. This time was distributed in a four-point dither optimised for extended sources. Each dither consisted of two integrations with 39 groups each. To support the background subtraction, a complementary MRS pointing at a nearby sky region was obtained using the same MRS configurations and observing setup as for SPT0311-58 but with only two dithers. The SHORT and MEDIUM spectral ranges cover eight independent spectral bands in the 4.9 to 28.1 $\mu$m spectral range (see \citealt{Wells+15, Labiano+21} for details).\footnote{Check this  \href{https://jwst-docs.stsci.edu/jwst-mid-infrared-instrument/miri-observing-modes/miri-medium-resolution-spectroscopy}{JDox page} for information about MRS channels and bands.}

\subsection{MIRIM F1000W calibration}
\label{sec:2.2.}

The MIRIM F1000W data were calibrated with version 1.6.2 of the JWST pipeline and context 0937 of the Calibration Reference Data System (CRDS). The CRDS context includes inflight dark current and flat-fields reference files. The reduction was done in an iterative way. We first ran the usual pipeline: (stage one) ramp to slopes, (stage two) photometric and astrometry calibration, and (stage three) drizzling, re-mapping, stacking, and mosaicking with default parameters. The mosaic resulting from this iteration presented a background gradient, with the bottom fifth of the field of view (FoV) darkening by up to $\sim$2\% compared to the (much flatter) rest of the image (far away from the SPT0311-58 source in any case). There were also vertical stripes at the $<$0.5 background level.

In order to subtract the background and remove the gradient and striping artefacts, we performed a second iteration of the reduction. We started this iteration by detecting objects in the mosaic using Sextractor and producing a segmentation map that we enlarge by 5 pixels to account for the low surface background regions of the galaxies. Next, we applied a column and row filtering of the data (subtracting medians of each column and then of each row), masking first the objects found in the first mosaic with a mathematical morphology dilation of 5 pixels. We also estimated the background in 128$\times$128 pixel$^2$ boxes and fitted a smooth surface to remove the gradient in the bottom part of the image (although the additive or multiplicative nature of this gradient is still not clear). We then obtained new intermediate calibrated files (i.e. the output of stage two of the pipeline) with flat null backgrounds, which we ran through stage three to produce a new mosaic. We then repeated the procedure once more to get better object masking and obtained final background-subtracted mosaics with pixel scales of 0.1109$\arcsec$ (the nominal pixel size) and 0.06$\arcsec$.

Another alternative method was used for the stripe removal and background homogenisation. After detecting and masking objects in the different dithered exposures, we calculated a background map to correct a given image. In the construction of the background map, we first removed the median of each image, then combined the images in a stacked background frame, and finally scaled the stack by the median background of the image we were interested in correcting. Although this method has been very successful in dealing with the background in some other datasets (e.g. PRIMER data, Pérez-González et al., in prep.), the limited number of dithering positions in the SPT0311-58 dataset did not allow for a correct background nor for artefact removal using this more advanced method.

The image with 0.06$\arcsec$ was used throughout the analysis, as it provides good matching with the ancillary high angular resolution (beam size of 0.07$-$0.08$\arcsec$) ALMA data \citep{Spilker+22}. The WCS of these mosaics were off by $\sim$0.5$\arcsec$ but were corrected when performing the additional astrometry corrections for comparison with ALMA and HST data (see $\S$\ref{sec:2.5.}). The  5$\sigma$ depth of the final mosaic reaches 25.5\,mag for a 0.32$\arcsec$ circular aperture (around our source), the radius being the full width at half maximum (FHWM) of the point spread function (PSF) in the F1000W band, for which we calculated a 30\% aperture correction, that is, the data reaches 25.2\,mag 5$\sigma$ depth for point-like sources.

\subsection{MRS calibration}
\label{sec:2.3.}  

The MRS observations were processed with version 1.7.3 of the JWST calibration pipeline and context 0977 of CRDS.\footnote{Check this  \href{https://jwst-docs.stsci.edu/jwst-science-calibration-pipeline-overview}{JDox page} for general information about the JWST calibration pipeline. Review the \href{https://jwst-docs.stsci.edu/jwst-mid-infrared-instrument/miri-features-and-caveats}{MIRI Features and Caveats} and \href{https://jwst-docs.stsci.edu/jwst-calibration-pipeline-caveats}{Pipeline Caveats} webpages for the latest status and information about the MIRI performance and calibration pipeline known issues.\label{Prueba_footnote}} In general, we followed the standard MRS pipeline procedure (\citealt{MRSpipeline, bushouse_howard_2022_6984366}; and \citealt{Alvarez-Marquez+19_mrs,Alvarez-Marquez+22} for examples of MRS data calibration) but included some variations that were needed to improve the quality of the final calibrated products for deep MRS observations. SPT0311-58 is a faint source whose main expected emission is around the Pa$\alpha$ and/or H$\alpha$ lines and is without detectable continuum emission. Therefore, the noise in the long integration ($\sim$7.5 ks) is dominated by the background and detector effects, and SPT0311-58 is not detected in individual exposures.   

The first stage of the MRS pipeline, which performs the detector-level corrections and transforms the ramps to slope detector products, was run skipping the reset step (Morrison et al. in prep.). Versions 1.6.2, or higher, of the pipeline includes the reset correction in the CRDS dark reference file. Detectors could have warm pixels, which are transient and likely due to the effect of previous cosmic ray events. For each MRS detector, we obtained the median of all background and on-source slope images, independent of the pointing. The derived median slope images were used to identify the warm pixels in each detector by applying a classical sigma clipping algorithm. The detected warm pixels were updated in the data quality array of the individual slope images. The first MRS integration of a visit presents a dark current ($\sim$25\%) lower than the following ones. We corrected for this effect in each of the slope images by subtracting the median values of all pixels located in the inter-channel area of the detector, which is usually not illuminated by the sky and/or telescope emission. We found that the standard pipeline identifies and corrects most of the cosmic ray (CR) events, but the so-called CR showers and/or highly energetic CRs still leave some relevant residual effects.$^{\ref{Prueba_footnote}}$ The residuals of the CR showers and additional detector effects are more relevant in channels one than three due to the differences in the background levels.

The second stage of the pipeline, which provides fully calibrated detector images, was run skipping the straylight and background pixel-by-pixel subtraction steps. Our MRS dataset is dominated by the background emission and/or detector effects, and therefore, the straylight correction provides only second-order corrections. Additionally, the straylight correction could provide over subtraction in some areas of the detector due to the presence of bright or hot pixels. The pixel-by-pixel background was performed as the last step of the second stage of the pipeline and after the photom/flat-field and flat fringe corrections were done. We generated a master background detector image for each MRS channel and band. This was done by obtaining the median, independent of the pointing, for all background and on-source fully calibrated detector images. The master background detector image was subtracted from each of the fully calibrated detector images. The decision to subtract the background at the end of the second stage was driven by the fact that the main source is not detected in individual exposures and the noise in the fully calibrated detector images is dominated by background or detector effects. Therefore, this pixel-by-pixel background subtraction would correct any fringe or detector level residuals as well as any features injected by the photometry and flat-field correction. 

The third stage of the pipeline, which combines all the exposures and provides the 3D spectral cubes, was run using default settings on the background subtracted and fully calibrated detector images. This process produced eight 3D spectral cubes, one for each band  (SHORT and MEDIUM) of the MRS channels, with spatial and spectral sampling of 0.13"~$\times$~0.13"~$\times$~0.001~$\mu$m for channel one, 0.17"~$\times$~0.17"~$\times$~0.002~$\mu$m for channel two, 0.20"~$\times$~0.20"~$\times$~0.003~$\mu$m for channel three, and 0.35"~$\times$~0.35"~$\times$~0.006~$\mu$m for channel four. The resolving power ranges from 4000, in channel one, to 1500, in channel four, correspond to a FWHM of 75km\,s$^{-1}$ to 200km\,s$^{-1}$ \citep{Labiano+21}. The WCS of the 3D spectral cubes were corrected using the MIRIM observations taken in simultaneous mode with the MRS (see $\S$\ref{sec:2.5.}). 

\subsection{ALMA and HST ancillary data}
\label{sec:2.4.}

\begin{figure*}
   \includegraphics[width=\linewidth]{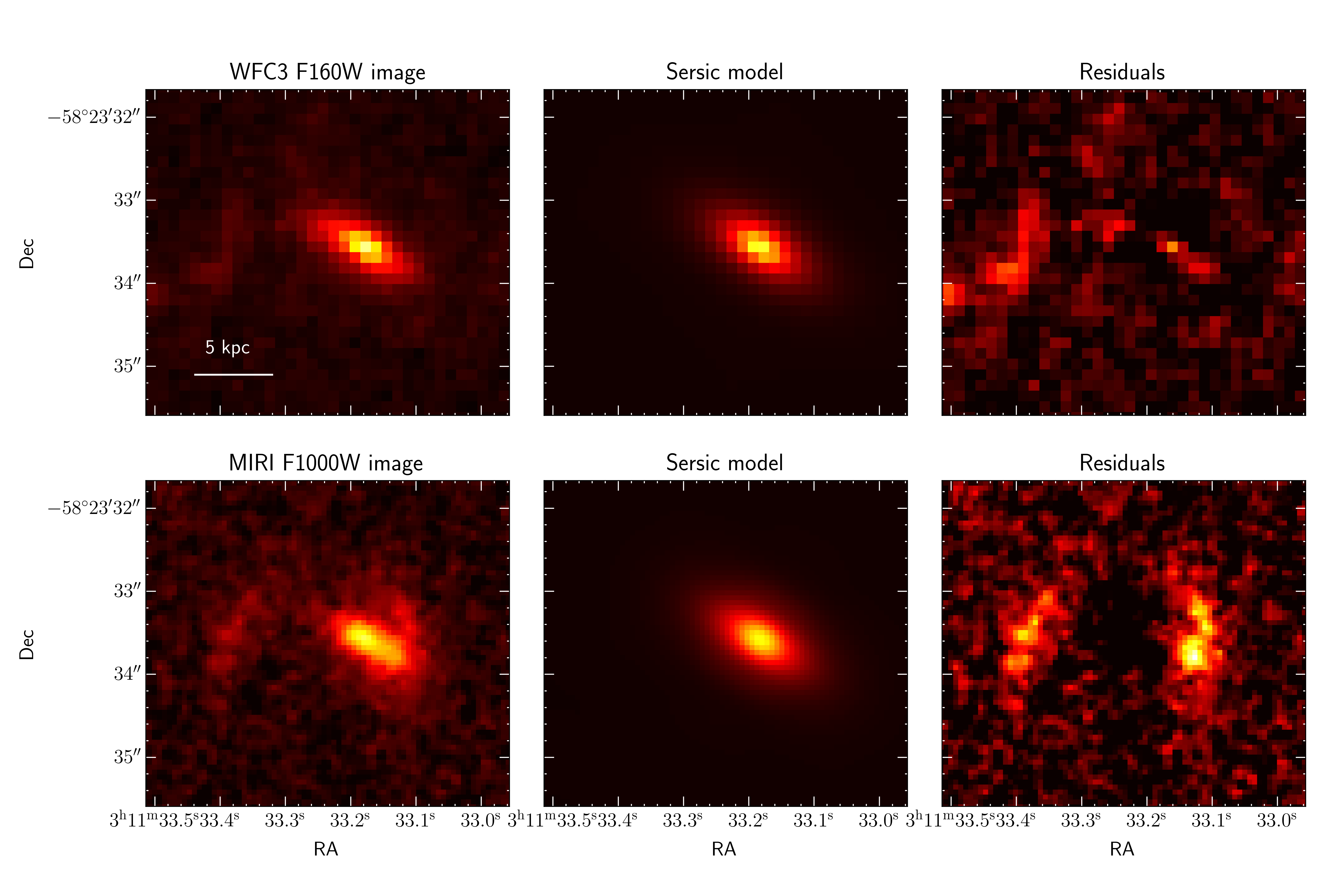}
      \caption{Lensing galaxy removal. The top (bottom) row of this figure shows, from left to right, the WFC3/F160W (MIRI/F1000W) image of SPT0311-58, the Sersic models from the lensing galaxy, and the residual images used for further analysis, respectively.}
         \label{fig:lens_remov}
\end{figure*}

The \CII\ observations were carried out with ALMA under program IDs (2016.1.01293.S and 2017.1.01423.S: PI Marrone). The data is described in detail in \citet{Spilker+22}, and we mostly followed the same reduction procedure as described there. A particular difference was that we applied one round of phase-only self-calibration to the data after running it through the ALMA pipeline. We imaged the rest-frame 160$\mu$m continuum (observed at $\sim$240\,GHz) using the line-free spectral channels by excluding a $\pm1500$\,km\,s$^{-1}$ window around the \CII\ emission line and using a Briggs weighting scheme with a robust parameter of 0.5, resulting in a synthesised beam of $0\farcs093$ x $0\farcs085$ and an rms noise of 8.0\,$\mu$Jy\,beam$^{-1}$. We also generated a continuum-subtracted data cube with a channel spacing of 30MHz ($\sim$37\,km\,s$^{-1}$), which has an rms noise of 0.10\,mJy\,beam$^{-1}$ per 30MHz channel. An integrated \CII\ image was generated from this data cube by summing up all emission within the channels that showed \CII\ emission. For visualisation purposes, we also generated an integrated \CII\ image from a continuum-subtracted data cube with a channel spacing of 150MHz where all spaxels below three times the rms noise were masked in order to highlight the clumps of \CII\ emission. The image is shown in Figure~\ref{fig:im_and_apertures}, although we note that all data analysis was performed on the non-masked image.

Complementary to the ALMA and MIRI data, ancillary near-infrared HST images of SPT0311-58 were retrieved from the Mikulski Archive for Space Telescope. Images were taken with the WPFC3 camera using the near-infrared filters F125W and F160W  (PI: Marrone, ID: 14740). These fully calibrated images with an integration time of $\sim$2800\,s for each filter have a pixel scale of $\sim$0.13\arcs and angular resolution FWHM$\sim$0.25\arcs, similar to that of our MIRI F1000W image.

\subsection{JWST-ALMA-HST astrometry}
\label{sec:2.5.}

To perform a proper comparison of the structures seen at different wavelengths, we needed to put all the different datasets into the same coordinate system. Since we were dealing with very high angular resolution data, this required a precision better than 100\,mas in the absolute positioning.
 
The absolute astrometry of the ALMA data was estimated from the position of a secondary source that was observed during the ALMA observations, which has a known position in the sky as tabulated in the ALMA source calibrator catalogue. Observations of this secondary source were interspersed between subsequent phase-target calibrator sequences. Analysis of the position of this secondary source yielded average positional deviations of less than 3 milliarcseconds, which even after accounting for the positional uncertainty in the source catalogue, yields absolute astrometric uncertainties on the SPT0311 observation smaller than 10 milliarcseconds.

The MIRI observations have position uncertainties due to the overall telescope pointing uncertainties (i.e. guide stars and roll angle). To correct for this effect, we used three stars from GAIA DR3 \citep{GaiaCollaboration+22} identified in the HST images, which yielded an uncertainty (1$\sigma$) in the final absolute positioning for the HST WFC3 images (i.e. F125W and F160W) of 37\,mas and 36\,mas in RA and Dec, respectively. The MIRI F1000W image was then aligned with the HST data by fitting a 2D Gaussian to the lensing galaxy, producing an uncertainty smaller than 30\,mas. Therefore, the astrometric uncertainties when comparing MIRI/F1000W structures with those identified in the ALMA and HST data corresponds to less than $\sim$50 and $\sim$30\,mas, respectively. 

The MIRIM observations were taken in a simultaneous mode with the MRS observations, although with an FoV adjacent to the MRS due to the relative offset in the sky between MIRIM and MRS. The MIRIM observations used the filters F770W and F1000W. We used these simultaneous MIRIM observations to align and correct the absolute astrometry of the MRS 3D spectral cubes. We replicated the approach followed in the case of the F1000W direct image and used the GAIA DR3 stars present in the FoV to correct the astrometry, yielding an uncertainty (1$\sigma$) smaller than half the MRS pixel-size (i.e. <90\,mas). 

\section{Photometry and spectroscopic measurements}
\label{sec:3.}
\subsection{Lensing Galaxy removal}
\label{sec:3.1.}

SPT0311-58 is a system lensed by a galaxy at $z$\,=\,1.43$\pm$0.36 \citep{Marrone+18} that is located at projected angular distances of $\sim$0.6\arcs and $\sim$1.6\arcs from the system's W and E galaxies, respectively (see left panel, Fig.~\ref{fig:lens_remov}). We removed the lensing galaxy to obtain an accurate image of the structure and photometry of the SPT0311-58 system. 

We used two independent tools, \texttt{Statmorph} \citep{Rodriguez-Gomez+19} and \texttt{GALFIT} \citep{Peng+02}, to model the emission of the lensing galaxy. Both tools are designed to fit the light profile of galaxies taking into account the PSF of the instruments. The lens model was created with the HST WFC3 F160W image (see $\S$\ref{sec:2.4.} for details), which has the advantage that the W component of SPT0311-58 is below the detection limit and does not affect the analysis. The PSF corresponding to this image is defined by TinyTim \citep{Krist+93}.

Robust values of ellipticity $\sim$0.60 and position angles of $\sim$60$\degr$ north-to-east for the major axis were found with both fitting codes. However, the Sersic index and the effective radius vary from 1.2 and 0.51\arcs (\texttt{Statmorph}) to 1.9 and 0.75\arcs (\texttt{GALFIT}). A complementary fit using \texttt{GALFIT} was performed based on the MIRIM F1000W image, and the PSF model was derived with WebbPSF \citep{Perrin+14}. The Sersic index and effective radius of this fitting in which the centre of the lensing galaxy was fixed to minimise possible biases caused by the presence of the W component are 1.98 and 0.84\arcs, respectively. The effective radius and Sersic index are correlated, and therefore the apparent differences in the parameters of the light profile are likely due to the compactness of the galaxy and the way \texttt{Statmorph} and \texttt{GALFIT} fit the inner structure with the combined PSF+galaxy light profile. These differences translate into an uncertainty of $<$5$\%$ in the apparent integrated flux of the lensing galaxy (i.e. 5.4$\pm$0.2\,$\mu$Jy).

To obtain a lens-removed MIRIM F1000W image of the SPT0311-58, we created independent models of the lensing galaxy based on the fitted light profiles for the F160W and F1000W images using \texttt{Statmorph} and \texttt{GALFIT}, as mentioned above. These models were then subtracted from the original MIRIM F1000W image, after a convolution with the MIRIM PSF and a flux normalisation, based on the average emission from the central $\sim$\,0.3\arcs radius of the lens galaxy. An example of this procedure can be seen in Figure~\ref{fig:lens_remov} where the modelling using the WFC3 F160W image is presented.

The different approaches to modelling the lensing galaxy yielded residual maps with a similar over-subtraction in the centre of the image ($\sim$12\% of the original flux) spatially coincident with the lensing galaxy. However, this effect, which seems to be independent of the method and image considered, has a minor impact on the photometry measurements of the SPT0311-58 clumps, especially on those outside of the lens galaxy major-axis. Since there is no preferred method, all the available lens-removed maps were taken into account during the photometry analysis (see $\S$\ref{sec:3.2.}). This way, the final  uncertainty in the flux measurements includes not only the noise properties of the data themselves but also the potential  uncertainties and/or biases associated with using a specific methodology in the subtraction of the lensing galaxy. Uncertainties in the photometry of the SPT0311-58 clumps (see $\S$\ref{sec:3.2.}) range from $\sim$5\% for clumps in the E component (e.g. EA to EE) as well as the WA clump in the W component to $\leq$15\% for the clumps closest to the lensing galaxy (e.g. WB, WC, WD). 

\subsection{MIRIM F1000W photometry}
\label{sec:3.2.}

\begin{figure*}
   \includegraphics[width=\linewidth]{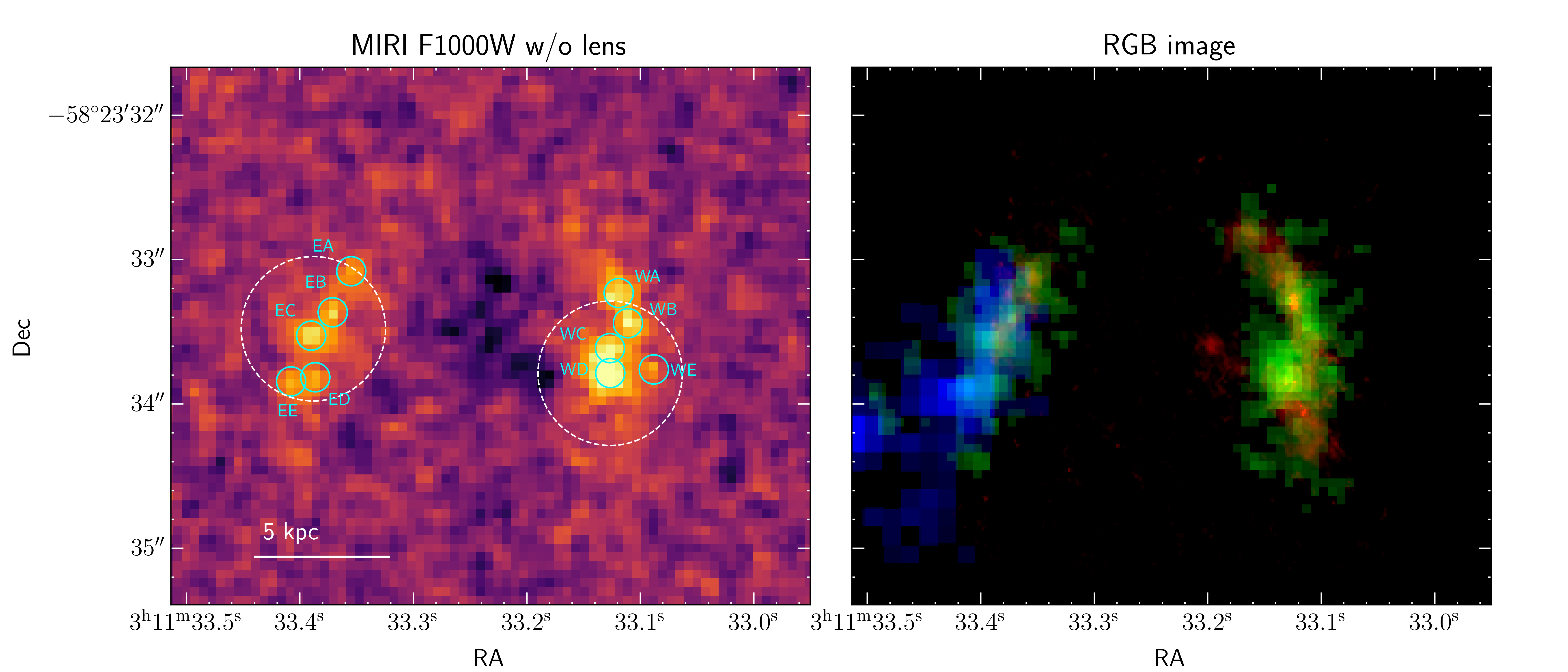}
      \caption{Imaging of the SPT0311-58 system. Left panel: MIRI F1000W image of SPT0311-58 system after removal of lensing galaxy. Cyan circles mark the position of the individual F1000W clumps, while white circles shows the MRS apertures used for the extraction of the Pa$\alpha$ spectra (see Sect.~\ref{sec:3.3.}). Right panel: RGB image where red, green, and blue colours show the ALMA \CII{}, MIRIM F1000W, and WPFC3 F160W lens-subtracted images representing the [\ion{C}{II}], rest-frame 1.25$\mu$m, and rest-frame 200nm continuum, respectively (see Sect.~\ref{sec:2.}).}
         \label{fig:MRS_app}
\end{figure*}

\begin{table*}
\caption{MIRI F1000W photometry derived for the MIRI/F1000W clumps.}          
\label{tab:F1000W_regions}      
\centering          
\begin{tabular}{c c c c c c}  
\hline\hline       
Region & RA & Dec & f$_\mathrm{obs}$ & f$_\mathrm{int}$ & L$_\mathrm{int}$\\ 
 & (J2000) & (J2000) & ($\mu$Jy) & ($\mu$Jy) & (10$^{9}$L$_\mathrm{\odot}$)\\ 
  (1) & (2) & (3) & (4) & (5) & (6) \\
\hline 
EA &  3h11m33.355s &  $-58\degr$23$'$33.083$\arcsec$ &  0.28$\pm$0.07 &  0.22$\pm$0.05 &  10.0$\pm$2.4 \\
EB &  3h11m33.371s &  $-58\degr$23$'$33.367$\arcsec$ &  0.30$\pm$0.05 &  0.23$\pm$0.04 &  10.5$\pm$1.8 \\
EC &  3h11m33.390s &  $-58\degr$23$'$33.529$\arcsec$ &  0.34$\pm$0.05 &  0.26$\pm$0.04 &  11.8$\pm$1.9 \\
ED &  3h11m33.386s &  $-58\degr$23$'$33.817$\arcsec$ &  0.22$\pm$0.04 &  0.17$\pm$0.03 &   7.8$\pm$1.5 \\
EE &  3h11m33.408s &  $-58\degr$23$'$33.846$\arcsec$ &  0.22$\pm$0.04 &  0.17$\pm$0.03 &   7.7$\pm$1.5 \\
WA &  3h11m33.119s &  $-58\degr$23$'$33.234$\arcsec$ &  0.39$\pm$0.12 &  0.18$\pm$0.06 &   8.1$\pm$2.5 \\
WB &  3h11m33.111s &  $-58\degr$23$'$33.443$\arcsec$ &  0.38$\pm$0.05 &  0.17$\pm$0.02 &   8.0$\pm$1.0 \\
WC &  3h11m33.127s &  $-58\degr$23$'$33.616$\arcsec$ &  0.50$\pm$0.06 &  0.23$\pm$0.03 &  10.3$\pm$1.3 \\
WD &  3h11m33.088s &  $-58\degr$23$'$33.763$\arcsec$ &  0.55$\pm$0.06 &  0.25$\pm$0.03 &  11.4$\pm$1.2 \\
\hline                  
\end{tabular}
\tablefoot{Column (1): Name of each region. Columns (2) and (3): Right ascension and declination for the centre of each aperture. Columns (4) and (5): Observed and intrinsic fluxes (i.e. corrected for lensing magnification factors) derived from an aperture of r$\sim$0.10\arcs. Column (6): Intrinsic F1000W luminosity.}
\end{table*}

The structure of the SPT0311-58 system appears to consist of a number of compact point sources (i.e. clumps). The photometry of each of the clumps was extracted in the clean 60\,mas F1000W image after the removal of the lensing galaxy. The energy enclosed within the FWHM (0.32\arcs) of the modelled PSF (WebbPSF) corresponds to $\sim$60\% to that at infinity (as also given by the aperture correction provided by the CRDS, i.e. \textit{jwst$\_$miri$\_$appcorr$\_$0008.fits}). The clumps detected in the F1000W image are typically separated by $\sim$0.2\arcs ($\sim$1\,kpc), and therefore, a direct measurement with an 0.16\arcs radius aperture would include emission from nearby clumps. To avoid the flux contamination within an aperture from close clumps, we used \texttt{GALFIT} to perform the photometry of each individual clump. All clumps were simultaneously fitted using the PSF derived from WebbPSF and considering each clump as an unresolved source centred in the emission peaks identified in the F1000W image (see Table~\ref{tab:F1000W_regions} and Fig.~\ref{fig:MRS_app}). To take into account the uncertainty induced by the lensing galaxy removal, we repeated the photometry for each of the different lens-subtracted models (see $\S$\ref{sec:3.1.}), defining the final observed fluxes as its average value. Associated errors were computed as the quadratic sum of the standard deviation of the models and the errors provided by GALFIT, which account for the errors in the background noise and contamination of clumps according to their relative positions.

Along with the photometry of the individual clumps, we also measured the integrated fluxes for the E and W galaxies. These values  were used as the observed 10$\mu$m photometry for the SED fitting (see $\S$\ref{sec:4.1.}). The observed (i.e. uncorrected by lens magnification) fluxes for the E and W galaxies are 1.6$\pm$0.3\,$\mu$Jy and 2.2$\pm$0.4\,$\mu$Jy, obtained by integrating over areas of $\sim$1.7 and 1.5\,arcsec$^2$. Aperture correction factors of $\sim$1.17 and 1.22 were applied for the E and W galaxies, respectively. These factors were calculated by simulating the clump distribution of each galaxy independently, taking each clump as a point source modelled with the WebbPSF model, and calculating the flux outside the areas. The aperture corrected fluxes calculated in mentioned apertures agree within less than 20\% with the total flux obtained by summing up all clumps. This result suggests that diffuse emission, if any, must represent a small fraction of the total measured flux.

\subsection{MRS Pa$\alpha$ and H$\alpha$ spectra}
\label{sec:3.3.}

\begin{figure*}[h]
   \includegraphics[width=\linewidth]{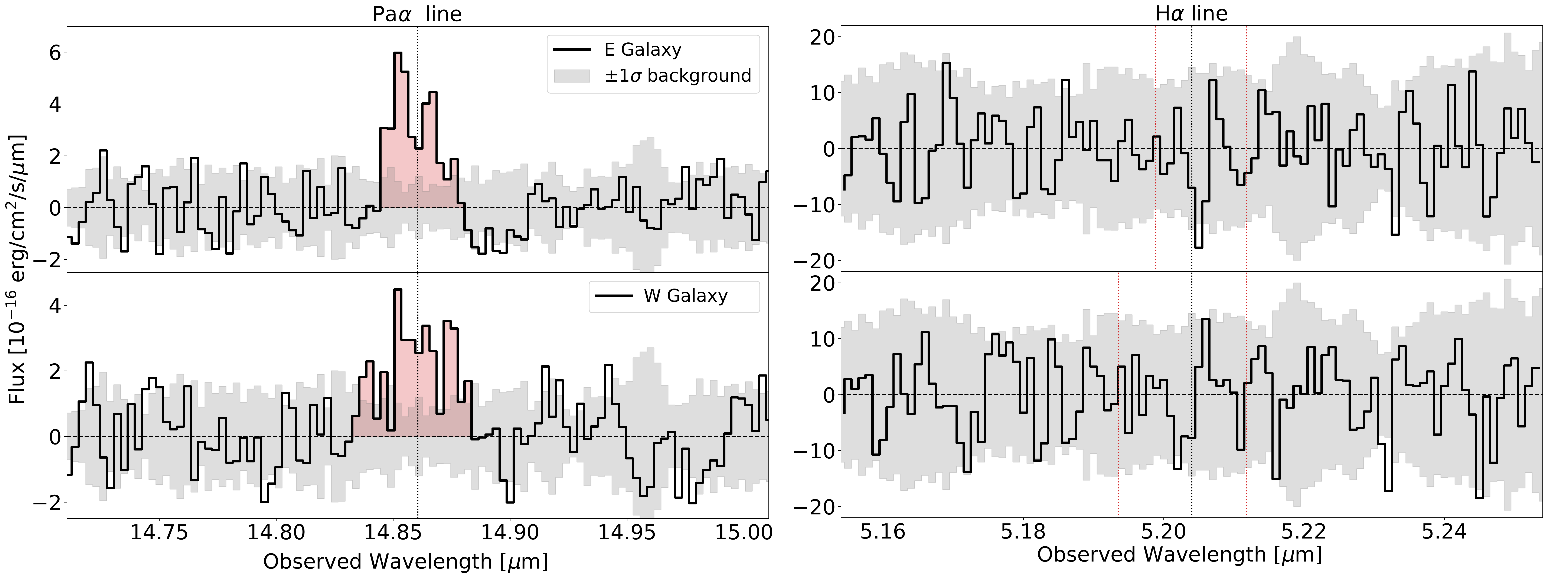}
      \caption{MRS spectra of Pa$\alpha$ (left plot) and H$\alpha$ (right plot) emission lines for  E (upper panel) and W (bottom panel) galaxies of SPT0311-58. Black lines: MRS extracted spectra. Red filled area: Wavelength integration range used to determine the Pa$\alpha$ line flux. Red dotted lines: Wavelength integration range used to determine the flux in Pa$\alpha$ line translated to the H$\alpha$ spectral range. Dotted black line: Central wavelength of Pa$\alpha$ line and expected central wavelength of H$\alpha$ line determined from Pa$\alpha$. Gray area: The $ \pm1\sigma$ calculated as the standard deviation form nine different background spectra.}
      
     \label{fig:Ha_spec}
\end{figure*}

The rest-frame H$\alpha$ and Pa$\alpha$ emission lines were redshifted to MRS channels 1SHORT and 3MEDIUM for SPT0311-58 at $z$=6.9. The Pa$\alpha$ line was detected in both E and W galaxies of SPT0311-58. Its emission appeared compact and concentrated around the main clumps identified in the MIRIM F1000W image,  EB+EC and WC+WD (see $\S$\ref{sec:4.2.} and $\S$\ref{sec:4.4.} for details), closely matching the ALMA [\ion{O}{III}]88$\mu$m emission \citep{Marrone+18}. However, within the current noise limits, H$\alpha$ was not detected in either of the two galaxies of the system. 

We performed 1D spectral extraction using standard aperture photometry \citep{larry_bradley_2022_6825092}. The Pa$\alpha$ spectra were extracted in an aperture radius equal to 0.5\arcsec ($\sim$2.7\,kpc) and centred at positions (RA[deg], Dec[deg]) equal to (47.8891167, $-$58.3926333) for the E galaxy and (47.888027, $-$58.392719) for the W galaxy (see white apertures in Figure \ref{fig:MRS_app}). With the same aperture, we also extracted nine 1D spectra in different positions of the MRS FoV clean of emission from the main galaxy system (lens galaxy + SPT0311-58). We combined the spectra to generate the 1D median and standard deviation spectra of the `background'. The median was subtracted from the Pa$\alpha$ spectra with the goal of removing any background residual left after the detector background subtraction was performed in the MRS calibration process (see Section~\ref{sec:2.3.}). Additionally, the Pa$\alpha$ spectra was affected by the continuum emission of the lens galaxy, and we subtracted it by fitting a linear function to the continuum surrounding the Pa$\alpha$ line. The final Pa$\alpha$ spectra is shown in Figure \ref{fig:Ha_spec} together with the 1$\sigma$ errors calculated from the standard deviation of the nine different background spectra. The Pa$\alpha$ line is spectrally resolved and centred at 14.860$\mu$m for both components. The flux of Pa$\alpha$ was calculated by integrating in a velocity range from $-$350\,km\,s$^{-1}$ to 450\,km\,s$^{-1}$ (E galaxy) and $-$650\,km\,s$^{-1}$ to 450\,km\,s$^{-1}$ (W galaxy) around its observed central wavelength (see red-filled area in Figure \ref{fig:Ha_spec}). These velocity ranges delimit the values at which the Pa$\alpha$ flux is positive. The Pa$\alpha$ line was detected with a significance of 9.7 and 7.8 and with observed fluxes of 1.07$\pm$0.11 and 1.09$\pm$0.14 $\times$10$^{-17}$\,erg\,s$^{-1}$\,cm$^{-2}$ for the E and W galaxies, respectively. The fluxes and spectra were corrected by aperture losses, assuming that the Pa$\alpha$ emission is a point source for the MRS angular resolution at 14.86$\mu$m (PSF FWHM = 0.6\arcs). The percentage of flux outside the selected aperture is 36.5\% of the total, which was found by using the latest MRS PSF models (Papatis et al., in prep.). 

The H$\alpha$ spectra were obtained following the same procedure as that for Pa$\alpha$ but with an aperture radius equal to 0.3\arcs to reduce the noise level. The H$\alpha$ line was not detected in either of the two galaxies (see Figure \ref{fig:Ha_spec}). The observed 3-sigma upper limits for the H$\alpha$ flux, assuming the same velocity interval as for Pa$\alpha$, are 7.0$\times$10$^{-18}$ and 8.6$\times$10$^{-18}$\,erg\,s$^{-1}$\,cm$^{-2}$ for the E and W galaxies, respectively.

\section{Results and discussion}
\label{sec:4.}

\subsection{The overall spectral energy distribution of SPT0311-58 galaxies: Star formation and internal extinction}
\label{sec:4.1.}

We performed an SED fitting analysis with CIGALE \citep{Burgarella2005,Noll2009,Boquien2019} for the E and W galaxies of SPT0311-58. The observed SED ranges from 1.25$\mu$m to 3\,mm, combining public ancillary photometry obtained with other facilities as well as the new MIRIM and MRS data points (see Table \ref{tab:Photo_SED} for details). The observed fluxes of Pa$\alpha$ and the 3$\sigma$ upper limits of H$\alpha$ were also used in the SED fitting analysis. The star formation history (SFH) was modelled using a constant star formation with ages ranging from 1 to 100\,Myr. We adopted the stellar population models from \cite{Bruzual&Charlot+03} with solar metallicity and the \cite{Chabrier+03} initial mass function. We included the nebular continuum and emission lines, using solar metallicity, electron density of 100\,cm$^{-3}$, an ionised parameter equal to $\log(U)$=$-$3, and the fraction of Lyman continuum photons escaping the galaxy set to zero due to absorption by the ISM. The assumption to fix the metallicity to solar, even for a system at z=6.9, was determined by the fact that SPT0311-58 is a massive (10$^{10-11}$\,M$_{\sun}$) and evolved system \citep{Tremonti2004,DeBreuck2019}. High-z starburst (e.g. \citealt{Doherty2020}) and/or ultra-luminous infrared galaxies (ULIRGs) galaxies (e.g. \citealt{Gracia-Carpio2011}) have shown extreme nebular conditions with electron densities larger than 100\,cm$^{-3}$ and $\log(U)$ in the range of -1 to -3. When we considered these extreme values in the SED fitting, we found  physical parameters that varied from what is presented in Table \ref{tab:prop_SED} by less than 5\% and within the uncertainties.

\begin{figure*}[h]
   \includegraphics[width=\linewidth]{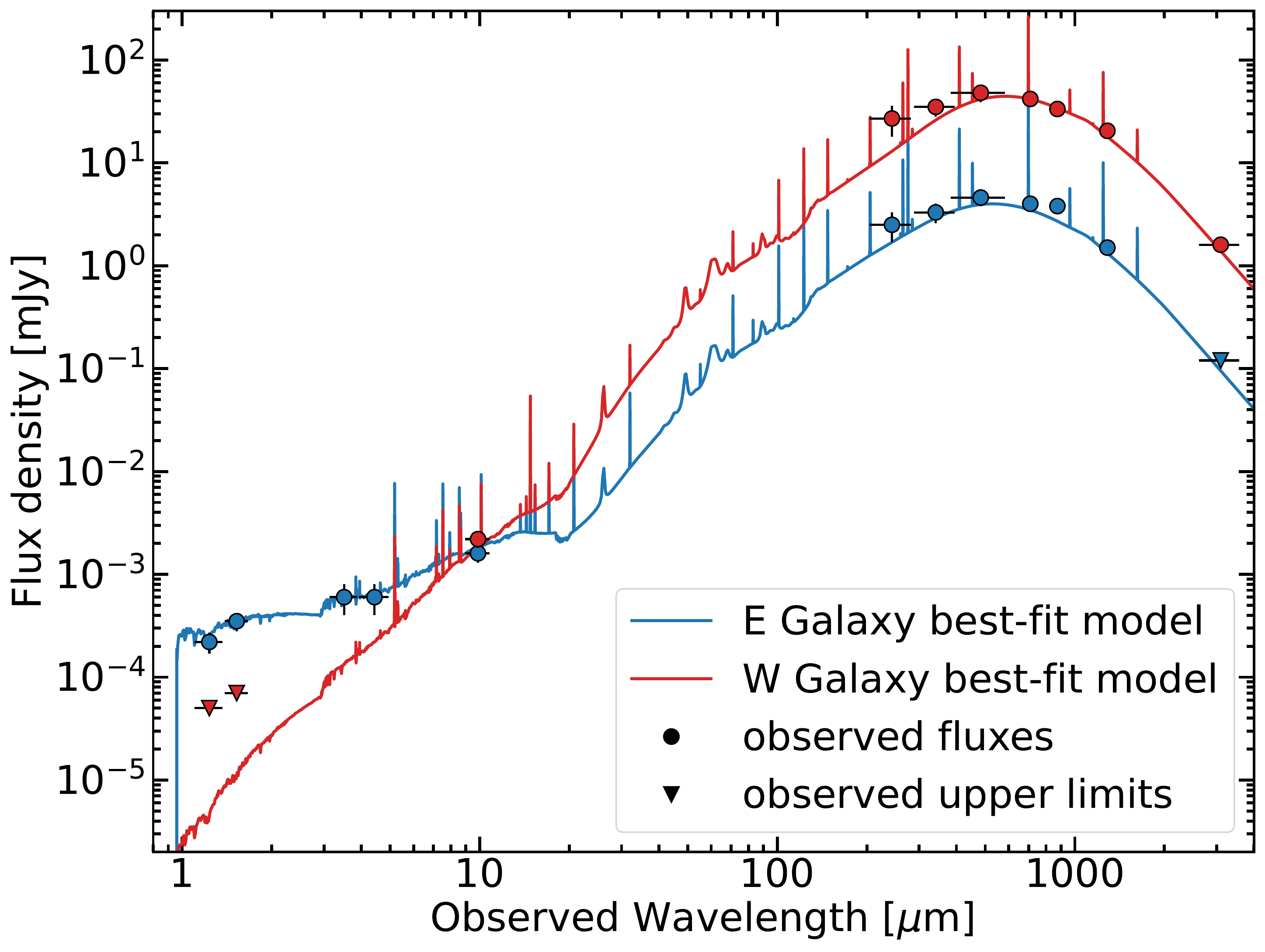}
      \caption{Observed near-infrared to millimeter SEDs and best-fit models derived from CIGALE SED fitting analysis of E and W galaxies. Blue line: Best-fit model for the E galaxy. Red line: Best-fit model for the W galaxy. Circles: Observed fluxes for the E and W galaxies. Triangles: Observed upper limits for the E and W galaxies. Observed quantities used in the fit are given in Table \ref{tab:Photo_SED}.}
     \label{fig:SED}
\end{figure*}

For the dust attenuation, we used the $dustatt\_modified\_CF00$ module of CIGALE \citep{Boquien2019}. It is based on the \cite{Charlot&Fall2000} dust attenuation model where the young stars (<10\,Myr) and the nebular emission located in the birth clouds (BC) suffer from additional attenuation compared with the stars (>10\,Myr) that have broken out and escaped into the ISM. This module is flexible and allows one to set up the slopes of the two components of the dust attenuation curve (BC and ISM), the $A_\mathrm{V}$ of the ISM component, and the factor that relates the $A_\mathrm{V}$ of the ISM and BC components. We left all parameters free in the SED fitting analysis, defining exactly the same initial parameters as \cite{LoFaro2017} in the analysis of the dust attenuation curve of ULIRGs at $z${}$\sim$2 (see the double power-law free case from Table 1 in \citealt{LoFaro2017}). The infrared emission was modelled using the \cite{Draine2007} dust models with the PAH mass fraction, the minimum radiation field, and the fraction of illumination from the minimum and maximum radiation fields parameters kept free. The SED fitting analysis for the E galaxy used all the discussed initial parameters, as it presents a relatively good wavelength coverage from rest-frame far-UV to far-infrared. In the case of the W galaxy, with its SED covering only the near- to far-infrared rest-frame, we fixed the initial parameters related to the slopes of the dust attenuation following the results of the E galaxy.               

\begin{table}[h]
\caption{Summary of the photometry and emission line fluxes used in the SED fitting analysis of the integrated E and W galaxies of SPT0311-58.}             
\label{tab:Photo_SED}      
\centering          
\begin{tabular}{lcc}  
\hline\hline       
Instrument/Filter & E galaxy & W galaxy \\ 
\hline                    
WPFC3/F125W & 0.22$\pm$0.01$\,\mu$Jy & <0.05$\,\mu$Jy  \\
WPFC3/F160W & 0.35$\pm$0.01$\,\mu$Jy & <0.07$\,\mu$Jy  \\
IRAC/3.6$\mu$m & 0.6$\pm$0.2$\,\mu$Jy & ---  \\
IRAC/4.5$\mu$m & 0.6$\pm$0.2$\,\mu$Jy & ---  \\
MIRIM/F1000W & 1.6$\pm$0.3$\,\mu$Jy & 2.2$\pm$0.4$\,\mu$Jy  \\
SPIRE/250$\mu$m & 2.5$\pm$0.8\,mJy &  27$\pm$9\,mJy \\
SPIRE/350$\mu$m & 3.3$\pm$0.7\,mJy & 35$\pm$7\,mJy  \\
SPIRE/500$\mu$m & 4.6$\pm$0.8\,mJy &  48$\pm$9\,mJy \\
ALMA/710$\mu$m & 4.0$\pm$0.3\,mJy &  41.8$\pm$0.7\,mJy \\
ALMA/869$\mu$m & 3.8$\pm$0.4\,mJy &  33.4$\pm$0.6\,mJy \\
ALMA/1.26mm & 1.5$\pm$0.1\,mJy & 20.5$\pm$0.4\,mJy  \\
ALMA/3mm & <0.12\,mJy &  1.60$\pm$0.05\,mJy \\
\hline 
MRS/H$\alpha$ & <0.70 & <0.86 \\
MRS/Pa$\alpha$ &  1.07$\pm$0.11 & 1.09$\pm$0.14 \\
\hline
\hline
\end{tabular}
\tablefoot{The WPFC3 and MIRI photometry were obtained using the same aperture (see Sections \ref{sec:2.4.} and \ref{sec:3.2.}). The IRAC, SPIRE, and ALMA photometry are from \cite{Marrone+18}. The H$\alpha$ and Pa$\alpha$ fluxes are in units of 10$^{-17}$\,erg\,s$^{-1}$\,cm$^{-2}$ (see Section~\ref{sec:3.3.}).}
\end{table}

The results of CIGALE SED fitting analysis for the two galaxies of the SPT0311-58 system are summarised in Table \ref{tab:prop_SED} and Figure \ref{fig:SED}. Table \ref{tab:prop_SED} presents the main physical and observed parameters derived from the CIGALE Bayesian analysis, and Figure \ref{fig:SED} shows the SED models that best fit the observed photometry and emission line fluxes. Assuming a constant SFH, the dominant stellar population is compatible with ages ranging from 15 to 35 Myrs in both components of SPT0311-58 and a total stellar mass equal to (1.4$\pm$0.4)$\,\times$10$^{10}$ and (8.0$\pm$2.4)$\,\times$10$^{10}$ M$_{\odot}$ for the E and W galaxies, respectively. A two component SFH combining an old (>100\,Myr) and young (<50\,Myr) stellar population was also tested. It yielded similar results for the young stellar population but with a probability of having an additional old (>100\,Myr) stellar population with a mass of up to 25\% of the young stellar population. Even if an ageing stellar population is expected, there is so far not enough information in the rest-frame far-UV to near-infrared SED to establish its contribution. Therefore, the SED fitting analysis uses the constant SFH that identifies a young stellar population as the one dominating the SED. 

The dust attenuation curve derived from the E galaxy is compatible with the one obtained for ULIRGs galaxies at $z$\,$\sim$\,2 \citep{LoFaro2017}. The derived dust attenuation curve is flatter than Calzetti's  \citep{Calzetti+00}, producing a relatively lower attenuation at far-UV wavelengths and a higher one at near-infrared wavelengths. Similar results have been found in galaxies with high stellar masses (>10$^{10}$\,M$_\odot$) and/or in the ULIRG range (e.g. \citealt{Alvarez-Marquez+19_SED}). We also tested the SED fit with a more standard dust attenuation curve (as used for the Pa$\alpha$/H$\alpha$ extinction discussed in Section \ref{sec:4.3.}), which combines the Calzetti's dust attenuation law \citep{Calzetti+00} for the stellar component and the Cardelli's Milky Way extinction law \citep{Cardelli+89} for the nebular emission. This approach failed to simultaneously fit the rest-frame stellar+nebular far-UV and far-infrared SED and the observed Pa$\alpha$ fluxes. This suggests the potential need of a different dust attenuation curve than that of Calzetti, which is the one commonly used at high-$z$. This result would require further analysis involving additional data that is beyond the scope of this paper.

The derived visual ($\lambda$\,=\,550\,nm) attenuation ($A_\mathrm{V}$) seen by the nebular emission and young (<10\,Myr) stellar population is 5.6$\pm$1.7 and 10$\pm$1 mag in the E and W galaxies, respectively. The $A_\mathrm{V}$ was reduced to 1.3$\pm$0.3 and 5.0$\pm$0.4 mag in the case of the older (>10\,Myr) stellar populations. We obtained an SFR equal to 740$\pm$100 and 3640$\pm$500\,M$_\mathrm{\odot}$\,yr$^{-1}$ that yielded an sSFR ($\equiv$\,SFR/$M_{*}$) of 53$\pm$22 and 45$\pm$20\,Gyr$^{-1}$ for the E and W galaxies, respectively. Finally, our SED fitting analysis was able to fit the observed photometry and derive H$\alpha$ and Pa$\alpha$ fluxes in agreement with the ones obtained from the MRS spectra. 

\begin{table}[h]
\caption{Summary of the physical properties and emission lines fluxes determined from the SED fitting analysis of the integrated E and W galaxies of SPT0311-58.}             
\label{tab:prop_SED}      
\centering          
\begin{tabular}{lccc}  
\hline\hline       
Parameter & Units & E galaxy & W galaxy\\ 
\hline     
Age Stars & Myr & 21$\pm$7 & 26$\pm$10 \\
$M_\mathrm{*}$ & $\times$10$^{10}$\,M$_{\odot}$ & 1.4$\pm$0.4 & 8.0$\pm$2.4  \\
SFR & M$_{\odot}$yr$^{-1}$ & 740$\pm$100 & 3640$\pm$500 \\
sSFR & Gyr$^{-1}$ & 53$\pm$22 & 45$\pm$20 \\
$A_\mathrm{V-ISM}$ & mag & 1.3$\pm$0.3 & 5.0$\pm$0.4 \\
$A_\mathrm{V-BC}$ & mag & 4.3$\pm$1.6 & 5.0$\pm$0.9 \\
$F_\mathrm{H\alpha}$ & $\times$10$^{-18}$ergs$^{-1}$cm$^{-2}$ & 5.2$\pm$2.2 &  1.2$\pm$0.5 \\
$F_\mathrm{Pa\beta}$ & $\times$10$^{-18}$ergs$^{-1}$cm$^{-2}$ & 2.3$\pm$0.4 & 1.6$\pm$0.4 \\
EW$_\mathrm{Pa\beta}$ & $10^{-3}$ $\mu$m (rest-frame) & 5$\pm$1 & 4$\pm$2 \\
$F_\mathrm{Pa\alpha}$ & $\times$10$^{-18}$ergs$^{-1}$cm$^{-2}$ & 8.8$\pm$1.3 & 10$\pm$2 \\
$L_\mathrm{IR}$ & $\times$10$^{12}$L$_{\odot}$ & 5.3$\pm$0.7 & 28$\pm$4 \\
\hline
\hline
\end{tabular}
\tablefoot{$A_\mathrm{V-BC}$ and $A_\mathrm{V-ISM}$ are the visual ($\lambda$\,=\,550\,nm) attenuation of the birth cloud and interstellar medium components of the dust attenuation curve, respectively. The nebular emission and young stars (<10\,Myr) are attenuated with an $A_\mathrm{V}$ equal to the sum of the $A_\mathrm{V-BC}$  and $A_\mathrm{V-ISM}$, and older stars (>10\,Myr) with an $A_\mathrm{V}$ equal to $A_\mathrm{V-ISM}$. $F_\mathrm{H\alpha}$, $F_\mathrm{Pa\beta}$, and $F_\mathrm{Pa\alpha}$ are the H$\alpha$, Pa$\beta$, and Pa$\alpha$ observed fluxes derived from the SED fitting analysis. The SFR, $M_{*}$, and $L_\mathrm{IR}$ were corrected for magnification.}
\end{table}

\subsection{The clumpy stellar structure of SPT0311}
\label{sec:4.2.}

The MIRI F1000W filter traces the near-infrared rest-frame (1.14$-$1.39\,$\mu$m) at the redshift ($z$=6.9) of SPT0311$-$58. Within this band, there are no strong emission lines, and the contribution of the nebular continuum is negligible. Therefore, the F1000W image traces the stellar light of this system.  As long as its equivalent width (EW) is large enough in comparison to the bandwidth of the filter (i.e.\,2\,$\mu$m), the Pa$\beta$ hydrogen line is the only line that could potentially contribute to the F1000W flux. However, the Pa$\beta$ equivalent width drops to values of less than 0.016\,$\mu$m for constant star formation older than 10\,Myr \citep{Leitherer+99}, which is in good agreement with the  Pa$\beta$ EW values derived from the SED fitting (see Table \ref{tab:prop_SED}). Only stellar populations dominated by extremely young stars with ages of less than 4\,Myr could have Pa$\beta$ equivalent widths closer to 0.1$\mu$m, contributing up to 40\% of the observed flux in the F1000W band at $z$=6.9. Similarly, the nebular continuum could also contribute significantly (i.e. above 20\%) to the total (stellar plus nebular) continuum emission at wavelengths of 1 to 2\,$\mu$m for stellar populations younger than 4\,Myr (\citealt{Reines+10}). However, the SED fitting confirms that a scenario where extreme young stellar populations dominate the F1000W light is unlikely. Thus, we considered that the effect of the nebular emission is negligible in the F1000W filter.

\begin{figure*}
    \includegraphics[width=\linewidth]{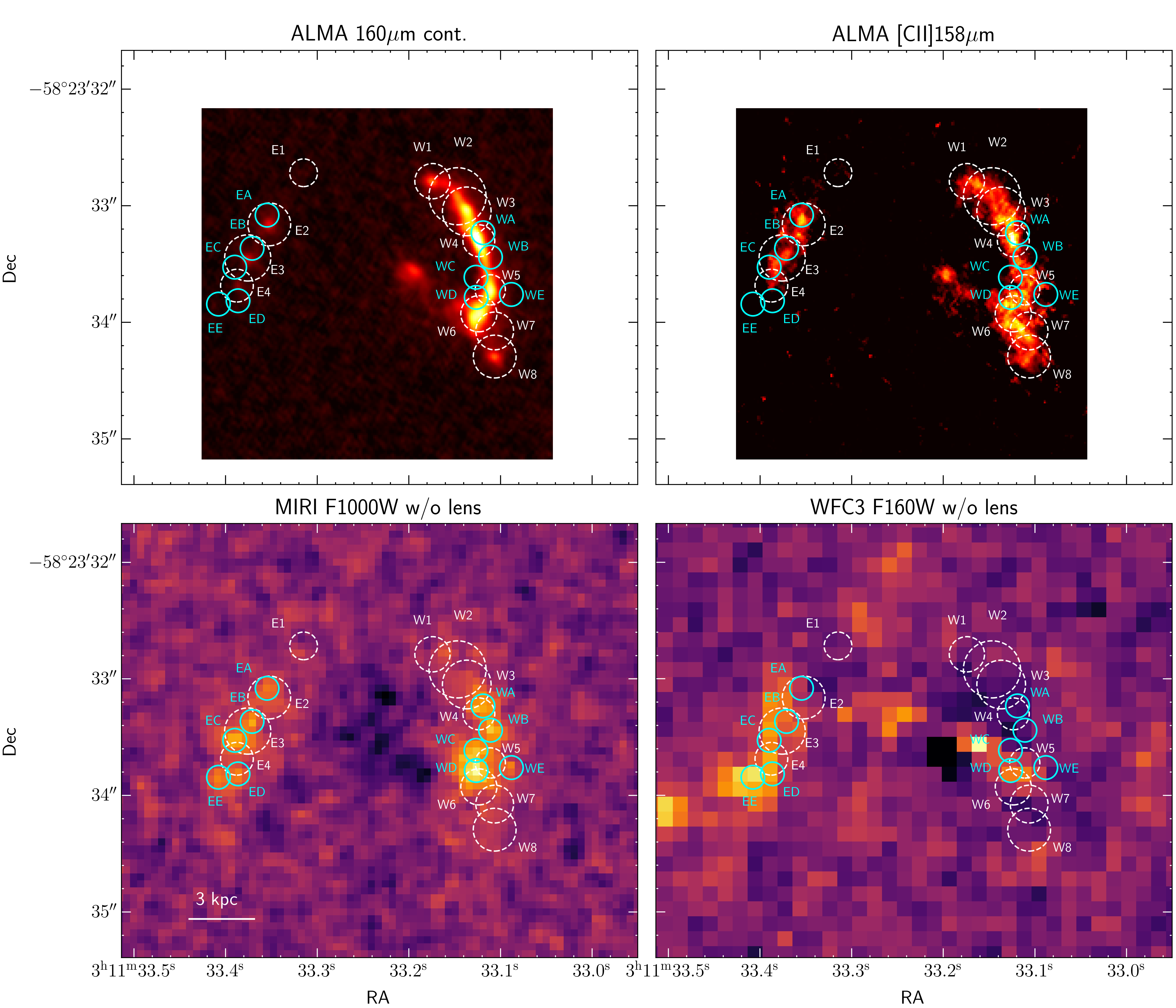}
      \caption{Spatial distribution of ALMA and MIRI clumps. From left to right, ALMA rest-frame 160$\mu$m continuum, ALMA [\ion{C}{II}]158\,$\mu$m, MIRI/F1000W, and WFC3/F160W images of SPT0311-58. The MIRI and HST images have been subtracted from the lensing galaxy (see $\S$\ref{sec:3.1.}). White dashed circles display the apertures used to extract the photometry centred on ALMA continuum clumps and defined by \citet{Spilker+22}. Cyan circles mark the clumps derived in this work from the F1000W image.}
         \label{fig:im_and_apertures}
\end{figure*}

The stellar light distribution of SPT0311-58 shows a clumpy structure in each galaxy of the system. The E galaxy consists of five clumps (EA to EE) distributed over a length of about one arcsec (i.e. $\sim$5.4\,kpc) with similar apparent fluxes and no one  clump dominating the light distribution. The W galaxy shows very similar characteristics, having four clumps (WA to WD, and an additional tentative WE) identified in a slightly more compact region of about 0.8\arcs (i.e. $\sim$4.3\,kpc) in length. The overall stellar structure is reminiscent of the 160$\mu$m continuum and [\ion{C}{II}]158$\mu$m line emission already known from high angular resolution ALMA imaging \citep[][see also \citealt{Marrone+18,Jarugula+21}]{Spilker+22}. However, differences are evident upon direct comparison, in particular when comparing with the 160$\mu$m continuum, as it shows a clumpier structure than the more diffuse [\ion{C}{II}] emission  (see Fig.~\ref{fig:im_and_apertures}). In general, the F1000W clumps appear to be somewhat spatially offset from the 160$\mu$m by 100 to 200 mas (i.e. $\sim$0.4$-$0.8 kpc at the redshift of the system). These offsets are at the 2 to 4 sigma level of the relative astrometry of ALMA and MIRI imaging, and therefore are considered real. However, it is difficult to conclude based just on the astrometry alone whether the clumps identified by the peak emission in the F1000W and 160$\mu$m ALMA images are physically the same or are two independent clumps. In the following paragraph, we take an approach where we consider that relative offsets of 100 mas (i.e. 2$\sigma$ positional uncertainties) are likely the same clump, while offsets of 150-200 mas can be considered as already independent.

In the E galaxy, clumps EA, EB, and EC appear to be associated with the 160$\mu$m clumps. While clump EA has a direct correspondence with the ALMA clump E2 identified in both the 160$\mu$m and [\ion{C}{II}] emission, clumps EB and EC are associated with the 160$\mu$m clump E3. However, both EB and EC are separated by $\sim$120\,mas (i.e. 0.65\,kpc) in opposite directions with respect to the 160$\mu$m continuum peak. A central dust lane bisecting the EB and EC clumps could explain this structure. The distribution seen in the [\ion{C}{II}] line reinforces this scenario, as the line emission is diffuse and extended over the region covered by clumps EB and EC. Nonetheless, F1000W clumps ED and EE do not have a counterpart in the ALMA images, in continuum, nor in the [\ion{C}{II}] distribution. These clumps, however, are detected in the HST F160W image (i.e. rest-frame UV light), with the clump EE being the strongest UV source of the two. These clumps can be interpreted as low-extinction regions with young high-mass stars emitting UV radiation and having a low gas and dust content. Finally, the very faint ALMA clump E1 was not detected in the F1000W image. 

In the W galaxy, the F1000W clumps appear to be more concentrated towards the bright central region of the far-infrared continuum and [\ion{C}{II}] emission. The 160$\mu$m clumps located in the most northern (W1, W2, and partially W3) and southern (W7 and W8) regions do not have a F1000W counterpart (see Figure~\ref{fig:im_and_apertures}).  The strongest far-infrared clump (W4) has a weak F1000W counterpart (WA). This suggests that the northern region of the W galaxy has a large amount of gas and dust that likely obscures the radiation in the near-infrared rest-frame. This condition was also independently confirmed by the non-detection of Pa$\alpha$ in the regions (see $\S$\ref{sec:3.3.}). The brightest F1000W clumps in the W galaxy (WB, WC, and WD) are offset by about 150$-$200\,mas (i.e. 0.8$-$1.1\,kpc) from the brightest 160$\mu$m emission clumps (W4 to W6). While WB and WC appear in regions with weak [\ion{C}{II}] emission, WD appears to be associated with a strong [\ion{C}{II}] emission that also appears offset from the continuum peak emission (W6). This suggests that the observed offsets between the emitting dust at 160$\mu$m, the [\ion{C}{II}] emitting gas, and the stellar light in the southern section of the W galaxy are partly due to the clumpy dust structure on (sub)kpc scales. Regions of high obscuration, as identified by the compact 160$\mu$m emission clumps, coexist with other regions that have much less dust and/or gas. In some discy low-$z$ luminos infrared galaxies (LIRGs), the cold molecular gas and stellar or ionised gas distribution show different clumpy structures on (sub)kpc scales \citep{Bellocchi+22}. Stellar clumps identified in the near-infrared continuum and emission lines are known to exist in low-$z$ LIRG discs \citep{Alonso-Herrero+06,Diaz-Santos+08}. 

\subsection{Internal extinction in the ionised interstellar medium and stars} 
\label{sec:4.3.}

The MRS data covers the two brightest hydrogen recombination lines in the optical (H$\alpha$) and near-infrared (Pa$\alpha$) range. The ratio of the upper limits in the observed flux for the H$\alpha$ with respect to that of the measured Pa$\alpha$ line allows one to derive a lower value to the internal extinction in both galaxies of the system. Assuming a H$\alpha$/Pa$\alpha$ recombination ratio of 8.6 (case B, for electron temperatures of 10$^4$\,K and electron densities of less than 10$^4$ cm$^{-3}$; \citealt{Hummer+87}), the internal nebular extinction is given as:

\begin{equation}
A_\mathrm{V}(\mathrm{nebular}) \geq   \frac{\log{8.6}-\log[F_\mathrm{obs}(\mathrm{H}\alpha)/F_\mathrm{obs}(\mathrm{Pa}\alpha)]}{0.4 \times [A(\mathrm{H}\alpha) - A(\mathrm{Pa}\alpha)]},
\end{equation}

\noindent
where $A$(H$\alpha$) and $A$(Pa$\alpha)$ are derived following Cardelli's Milky Way extinction law \citep{Cardelli+89}. The lower limits of the nebular visual extinction correspond to an $A_\mathrm{V}$ of 4.2 and 3.9 magnitudes for the E and W galaxies. 
These high internal nebular extinction values are similar to those measured in low-$z$ luminous (median $A_\mathrm{V}$ of 5.3) and ultra-luminous (median $A_\mathrm{V}$ of 6.5) galaxies based on the line ratios of near-infrared hydrogen lines \citep{Piqueras-Lopez+13,Piqueras-Lopez+16} or optical and near-infrared (H$\alpha$/Pa$\alpha$, H$\alpha$/Pa$\beta$; \citealt{Alonso-Herrero+06, Gimenez-Arteaga+22}). 
Also, intermediate redshift (median $z$= 1.55)  dusty star-forming galaxies do show large extinctions ($A_\mathrm{V}$ = 5.0$\pm$0.4, \citealt{Casey+17}). To derive the visual extinction for the stellar component, we follow Calzetti's prescription \citep{Calzetti+00} where E(B$-$V)(stellar) = 0.44$\times$E(B$-$V)(nebular) with $R_\mathrm{V}$ set as 4.05 \citep{Calzetti+00} and 3.1 \citep{Cardelli+89} for the stellar continuum and nebular lines. The lower limits to the visual extinction in the stellar component therefore correspond to 2.4 and 2.2 for the E and W galaxies, respectively. Internal extinction towards the stars has been previously estimated based on SED fitting (2.7$\pm$0.2 and $\leq$6 for E and W components, \citealt{Marrone+18}). The lower limits derived from the ratio of the hydrogen lines indicate an extinction for the stars in the E galaxy that is in good agreement with the SED value, while the SED fitting for the W galaxy provides an upper limit with a larger extinction value.  

The new SED fitting that includes the nebular continuum emission as well as the new F1000W and MRS data (i.e. observed 10$\mu$m continuum flux, Pa$\alpha$ emission line fluxes, and H$\alpha$ upper limits) provides stronger constraints and new independent values for the average extinction in each of the galaxies (see $\S$\ref{sec:4.1.}). When comparing with the nebular extinction derived from the recombination lines, one must bear in mind that the Pa$\alpha$/H$\alpha$ ratio provides the nebular extinction only towards the regions where the F1000W clumps are detected. In contrast, the SED provides an average of the internal extinction over the entire E and W galaxies. This average will likely show bias towards higher extinctions if a patchy dust distribution with strong extinction gradients on (sub)kpc scales is present. These distributions are typical in low-$z$ luminous and ultra-luminous galaxies \citep{Scoville+98, Scoville+00, Alonso-Herrero+06, Piqueras-Lopez+16, Gimenez-Arteaga+22}.
For the E galaxy, we derived extinctions $A_\mathrm{V}$(nebular)= 5.6 $\pm$ 1.7 for the nebular emission and young (i.e. less than 10 Myr) ionising stellar population and $A_\mathrm{V}$(stellar) = 1.3 $\pm$0.3 for older stellar populations. Much higher extinctions were obtained for the W galaxy, with  A$_\mathrm{V}$(nebular)= 10.0$\pm$0.9 and $A_\mathrm{V}$(stars) = 5.0$\pm$0.6, respectively. 

The derived nebular and stellar extinctions  assume a foreground dust screen with the ionised gas and young stars behind an additional slab of dust. However, the dust can be mixed with the young stars and ionised gas, especially in ultra-luminous infrared galaxies with large amounts of gas and dust such as the SPT0311-58 system. Under the assumption of mixed dust and gas geometry, the Pa$\alpha$/H$\alpha$ ratio reaches an asymptotic value of 1.55 \citep[see e.g.][]{Calabro+18} for large extinctions. The observed Pa$\alpha$/H$\alpha$ with lower values of 1.52 and 1.26 for the E and W galaxies already indicates very large extinctions that for this particular line ratio corresponds to $A_\mathrm{V}$ larger than six magnitudes, which is in agreement with the previous derivations. 

In summary, independent estimates of the extinction in the two galaxies that form the SPT0311$-$58 system indicate that 1) all methods provide similar estimates of the extinction within the observational uncertainties and assumptions; 2) the nebular extinction is extremely high in both systems and equivalent to no less than about four to five magnitudes in the visual; 3) in the foreground screen geometry, the stellar populations detected in the F1000W clumps have on average an extinction half that of the ionised gas, as derived from the SED fit; and 4) the amount of extinction is similar to that already known in low- and intermediate-$z$ dusty, infrared-luminous galaxies. 

\subsection{Kinematics of the ISM}
\label{sec:4.4.}

The ALMA data have shown a detailed picture of the gas kinematics traced by the [\ion{C}{II}]158$\mu$m and [\ion{O}{III}]88$\mu$m lines \citep{Marrone+18,Spilker+22}. Based on the integrated [\ion{C}{II}] emission, the redshift of the system is established at 6.9 (i.e. close to the centroid of the W galaxy of the system), with the E galaxy showing an apparent relative velocity of $+$700 km s$^{-1}$. The [\ion{C}{II}] kinematics cover a wide range in velocity, [$-$500,$+$500] km s$^{-1}$ (W) and [$+$500,$+$1200] km s$^{-1}$ (E), associated with the [\ion{C}{II}] clumps identified in the system \citep{Spilker+22}. The [\ion{O}{III}] light distribution shows a different spatial distribution than the [\ion{C}{II}] and is mostly concentrated in the E galaxy and the southern region of the W galaxy \citep{Marrone+18}. As a result, the velocity range of the [\ion{O}{III}] emitting gas is narrower than for [\ion{C}{II}], lacking  velocities in the [$-$500,$-$200] km s$^{-1}$ range associated with the northern [\ion{C}{II}] emission in the W galaxy. Other gas tracers such as the molecular CO(7-6) and the neutral [CI](2-1) lines show an even wider velocity range in the W component, with relative velocities up to $-$1500 km s$^{-1}$, likely associated with gas in the northern region of this galaxy. All these results show the complex kinematics associated with the different phases of the ISM. The Pa$\alpha$ line traces the kinematics of the fully ionised gas and therefore provides additional information on the kinematics of the ISM in the system. 

\begin{figure*}
    \includegraphics[width=\linewidth]{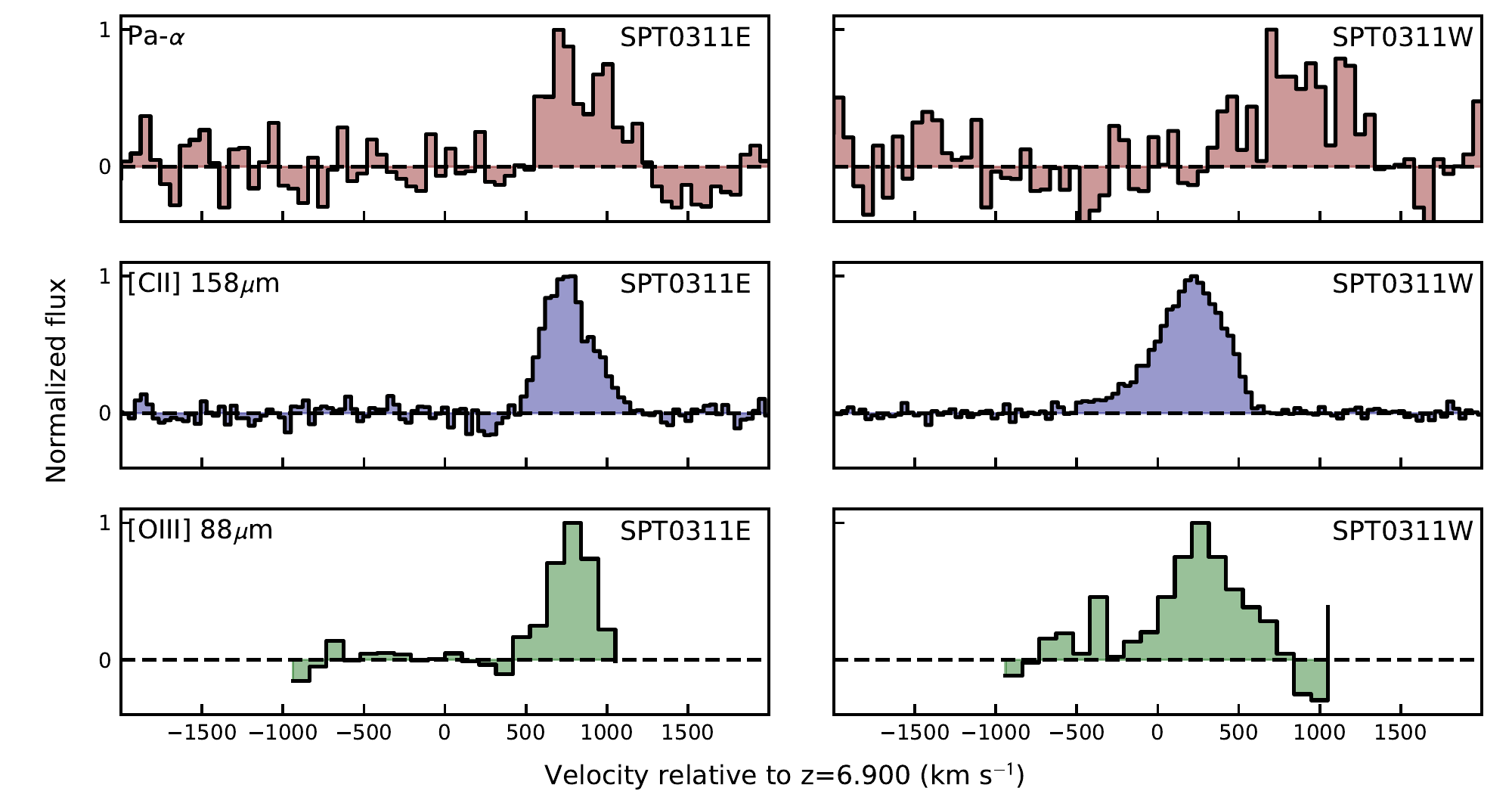}
     \caption{Figure showing the Pa$\alpha$ line profile for the E and W components of the system together with  [\ion{C}{II}]158$\mu$m and [\ion{O}{III}]88$\mu$m extracted using  same apertures as for  Pa$\alpha$ line, displayed on Fig.~\ref{fig:MRS_app}. The line profiles are normalised to the peak of the lines. }
        \label{fig:MRSvsALMA_lines}
\end{figure*}

The Pa$\alpha$ emission was identified in two well-defined regions. The strongest is located in the E galaxy with additional emission in the southern region of the W galaxy, while lacking detection in the galaxy's northern region. This distribution agrees with the position of the regions emitting the highly ionised gas traced by the [\ion{O}{III}]88$\mu$m line. 
The Pa$\alpha$ velocity peak in the E galaxy coincides with that of the [\ion{C}{II}] and [\ion{O}{III}] lines (Figure
~\ref{fig:MRSvsALMA_lines}), also showing an extension towards high redshifted velocities (up to 1200 km s$^{-1}$), that is, covering a velocity range similar to that of the [\ion{C}{II}] emission. According to the high angular resolution [\ion{C}{II}] imaging \citep{Spilker+22}, the velocity of the peak emission in the Pa$\alpha$ appears to be the same as the [\ion{C}{II}] clumps E3 and E4. The [\ion{O}{III}] emission, showing a narrower line profile than the Pa$\alpha$ line, appears to also be concentrated in clumps E3 and E4 \citep{Marrone+18}. 
However, the redshifted [\ion{C}{II}] velocities are associated with clump E2, which is located north of E3 and E4. Thus, in this galaxy the Pa$\alpha$ is tracing both the highly ionised gas as well as the more extended emission associated with additional [\ion{C}{II}] emitting gas. While the Pa$\alpha$ clumps cannot be resolved spatially with the MRS data, F1000W clumps EA to EC appear to be sharing the same kinematics and position (see Section~\ref{sec:4.2.}) as the main [\ion{C}{II}] clumps in the E galaxy (E2 to E4, see Figure~\ref{fig:im_and_apertures}). This indicates a physical connection between the ALMA and F1000 stellar clumps.  

The Pa$\alpha$ emission in the W galaxy is identified towards clumps WB$-$WD (see Figure~\ref{fig:MRS_app}), and there is no evidence of emission towards the northern region of the galaxy. These F1000W clumps have a spatial correspondence mainly with [\ion{C}{II}] clumps W5$-$W6 (see Figure~\ref{fig:im_and_apertures}).  The velocity range in the Pa$\alpha$ ionised gas covers a maximum range of about 1000 km s$^{-1}$ from about +300 to 1300 km s$^{-1}$. This range in velocity is similar to that observed in [\ion{C}{II}] and [\ion{O}{III}] (Fig. \ref{fig:MRSvsALMA_lines}) but with an offset of $+$700 km\,s$^{-1}$. There is no evidence of Pa$\alpha$ emission at the velocities traced by the peak emission of the [\ion{C}{II}] and [\ion{O}{III}] emitting gas. While there is certain evidence of velocities up to 700 km\,s$^{–1}$ in the highly ionised [\ion{O}{III}] line, neither [\ion{C}{II}] nor [\ion{O}{III}] show emission at velocities above this limit. Thus, if the velocity differences in the line profile between Pa$\alpha$ and the far-infrared lines in the W galaxy are real, these could indicate the presence of an unobscured ionised gas component with high velocities, with respect to the systemic velocity of the galaxy. Yet, due to the low SNR of the Pa$\alpha$ line profile, we cannot completely reject the possibility of residual (or unknown) effects in the calibration. Additional data with a better SNR are required before establishing the reality of this velocity offset in the W galaxy. Also, deeper ALMA spectroscopy for the [\ion{O}{III}]88$\mu$m line would help establish the presence of highly ionised gas at high velocities ($\geq$ 700 km s$^{-1}$) in the W galaxy. 

\subsection{Unobscured ongoing star formation}
\label{sec:4.5.}

The Pa$\alpha$ line is a tracer of young ($<$\,10\,Myr old) massive stars, and therefore, its luminosity was used to derive the rate of ongoing star formation in the system, after correcting for internal extinction. Since we were only able to measure a lower limit for the nebular extinction (see Section~\ref{sec:4.3.}), this SFR has to be considered as a lower limit to the total ongoing star formation. For a Kroupa IMF, the SFR can be given by the expression
\begin{equation}
\mathrm{SFR(M_\mathrm{\odot}\,yr}^{-1}) = 4.5\times 10^{-41}\times L(\mathrm{Pa}\alpha, \mathrm{ erg\,s}^{-1}) \times \Omega_\mathrm{ext}^{-1} \times \Omega_\mathrm{heat}^{-1}, 
\end{equation}

\noindent
where factors $\Omega_\mathrm{ext}$ and $\Omega_\mathrm{heat}$ represent the factors (0 $\leq$ $\Omega$ $\leq$ 1) associated with the internal extinction ($\Omega_\mathrm{ext}$) and the fraction of the ionising photons that do not ionise the ISM but heat the dust instead ($\Omega_\mathrm{heat}$).

Without internal extinction correction, the intrinsic (i.e. magnification corrected) Pa$\alpha$ luminosity corresponds to 213$\pm$22 and 128$\pm$16\,M$_\mathrm{\odot}$\,yr$^{-1}$ for the E and W galaxies, respectively. These values represent only a small fraction (28\% and 7\%) of the SFR derived from the [\ion{C}{II}] luminosities in the E and W clumps, $\sim$770 and 1760\,M$_\mathrm{\odot}$\,yr$^{-1},$ respectively \citep{DeLooze2014,Spilker+22}. Similar as well as even lower fractions can be obtained if the SFRs are derived from the integrated SED (740$\pm$100 and 3640$\pm$500\,M$_\mathrm{\odot}$\,yr$^{-1}$, see $\S$\ref{sec:4.1.}) or the infrared luminosity (540$\pm$175 and 2900$\pm$1800\,M$_\mathrm{\odot}$\,yr$^{-1}$, \citealt{Marrone+18}). It is worth pointing out that for the E galaxy, the SFR values based on the [\ion{C}{II}] line, the integrated SED, and the infrared luminosity are in very close agreement, in particular those derived using the [\ion{C}{II}] line and the SED fitting. In the W component, the high SFR values cover a much wider range, but they are still in agreement within the large uncertainties. The SFR based on [\ion{C}{II}] estimations has been seen to be lower than the ones derived from the infrared luminosity in starburst galaxies due to the so-called [\ion{C}{II}]-deficit (e.g. \citealt{Herrera-Camus2018}).

The factor to be applied to the Pa$\alpha$ extinction corrections for the E galaxy ranges from $\geq$1.8 to 4.4 based on the (Pa$\alpha$/H$\alpha$) ratio and SED fit, respectively. When correcting for these factors, the intrinsic (demagnified) SFR(Pa$\alpha$) is in the 383-937 M$_\mathrm{\odot}$\,yr$^{-1}$ range (i.e. between 50\% and 100\% of the SFR derived using far-infrared tracers).  This ratio is similar to those measured in low-$z$ LIRGs when comparing the extinction-corrected SFR derived from near-infrared lines with the total SFR derived in the far-infrared \citep{Alonso-Herrero+06, Piqueras-Lopez+16, Gimenez-Arteaga+22}. This ratio can be explained by a dominant fraction of the ionising photons that ionise the gas and subsequently becoming absorbed by the surrounding dust, instead of being directly absorbed by dust.

The situation for the W galaxy is different. The lower limit for SFR(Pa$\alpha$) (=230\,M$_\mathrm{\odot}$\,yr$^{-1}$), derived using the Pa$\alpha$/H$\alpha$ ratio, still represents a very small fraction of the SFR derived in the far-infrared (6$\%$ to 13$\%$ if the SED or infrared luminosity are considered). This is in the range measured in low-$z$ ULIRGs with SFR$_\mathrm{IR}>$100\,M$_{\odot}$\,yr$^{-1}$. In these galaxies, the extinction-corrected SFR(Pa$\beta$) or SFR(Pa$\alpha$) is a factor of four to ten times lower than the SFR values derived from their infrared luminosities \citep{Piqueras-Lopez+16,Gimenez-Arteaga+22}. To recover the SFR predicted by the far-infrared tracers, we applied the high extinction derived from the SED fit ($A_\mathrm{V}$(tot)=10$\pm$1). Assuming this extinction and the attenuation curve for SPT0311-58 (in agreement with the one measured for $z\sim$2 ULIRGs, \citealt{LoFaro2017}), the extinction-corrected SFR(Pa$\alpha$) is $\sim$4200$\pm$500\,M$_\mathrm{\odot}$\,yr$^{-1}$, which is just above the predicted SFR from the SED fit.

There are several effects that could explain the differences between the Pa$\alpha$/H$\alpha$ extinction-corrected Pa$\alpha$ values and those derived using far-infrared tracers. Recombination lines trace only very recent star formation during the last 10 Myr, while far-infrared radiation traces a wider range of stellar ages up to about 100 Myr. So part of the difference could be due to evolution of the star formation history over the last 100 Myrs.  To establish accurate ages for the stellar population, a full coverage of the rest-frame UV to near-infrared SED is needed, but this is not available so far for this galaxy. The second effect is the intrinsic clumpy and patchy nature of the dust distribution on (sub)kpc scales in dusty luminous infrared galaxies. This effect is well known at low-$z$ where two-dimensional dust structure showing large variations in the equivalent $A_\mathrm{V}$ have been measured in (ultra-)luminous infrared galaxies \citep{Scoville+98,Scoville+00,Piqueras-Lopez+13}. This characteristic could be playing a major role in the W galaxy, where on average very high extinctions are derived from the SED fit. In fact, the distribution of the ionised gas and 10$\mu$m emission in this galaxy is very different from that detected in both the 160$\mu$m continuum and [\ion{C}{II}] line. The Pa$\alpha$ and 10$\mu$m emission is concentrated in the central regions without evidence of emission in the northern region of the galaxy. This could suggest the star-forming clumps in the latter region are extremely obscured or that there is a lack of (young) stars able to ionise the surrounding medium. In contrast, the E galaxy shows a less extreme scenario. In this galaxy both the 10$\mu$m and Pa$\alpha$ emission show a distribution similar to that detected in the far-infrared. Additional regions detected in the rest-frame UV and near-infrared continuum are not detected in the far-infrared, pointing to regions with young stellar populations that are almost free of dust. A third effect in explaining the Pa$\alpha$ and far-infrared SFR could be associated with the direct absorption of Lyman continuum photons by the dust. In this case, recombination lines will not be produced while the ionising photons directly increase the flux at infrared wavelengths.

\subsection{Stellar to dynamical and gas mass ratios in the clumps}
\label{sec:4.6.}

The F1000W MIRI filter is a broad-band filter centred at 10$\mu$m with a width of 2$\mu$m. Therefore, F1000W traces the near-infrared rest-frame in SPT0311$-$58, covering 1.14 to 1.39$\mu$m, close to the standard ground-based photometric J-band. 
The stellar mass in each of the clumps was derived from the lensing-corrected F1000W observed flux,  assuming a mass-to-light (M/L) ratio for stellar populations with a range of ages (Table~\ref{tab:Mass_F1000W_reg}). In the following paragraphs we refer to the J-band luminosity and mass-to-light ratios as M/L for simplicity.

To derive the M/L, the F1000W filter transmission was convolved with redshifted spectral energy distribution of stellar populations with solar metallicity and constant SFRs over different periods of time ranging from 10 Myr up to 100 Myr. This range of ages was selected as the different tracers (Pa$\alpha$, [\ion{C}{II}], and overall SED) indicate the presence of stars with ages within this range. Some cosmological simulations of less massive high-$z$ galaxies \citep{Ceverino+18} have shown that the star formation history of these galaxies is complex, experiencing several bursts during periods of about 100 Myr at redshifts 6 to 15 (i.e. first 1 Gyr). Therefore, stellar populations with older ages (hundreds of Myr) should also be present in SPT0311-58. However, there are no observational constraints available to estimate their mass contribution in SPT0311-58. In the following analysis, stars older than 100 Myr are not considered. If such stars were contributing a significant fraction of the F1000W light, the M/L would be higher than that of the younger populations, and therefore the stellar mass derived from the F1000W flux would be higher by a large factor (e.g. M/L[500\,Myr]\,>\,6\,M/L[100\,Myr]). 

To establish the average M/L for different stellar populations, the mass-to-light ratios were derived using different spectral energy distributions and computed using different models (STARBURST99, \citealt{Leitherer+99}; BPASS, \citealt{Eldridge+17}; and Bruzual-Charlot, \citealt{Bruzual&Charlot+03}). We considered a Kroupa IMF \citep{Kroupa+01} for the STARBURST99 and BPASS models and a Chabrier IMF \citep{Chabrier+03} for the Bruzual-Charlot ones. Offsets between Chabrier and Kroupa IMFs are usually considered negligible (see \citealt{Madau&Dickinson+14}), and therefore no correction was applied during this analysis. The possible contribution of the nebular emission to the M/L has been studied using the STARBURST99 models. The nebular contribution represents 5\% of the total flux for the 10\,Myr continuum SFR, while smaller fractions were obtained for longer-lived constant SFRs. Near-infrared rest-frame M/L ratios and uncertainties were computed as the average value and standard deviation among the different models for each stellar population considered. Mass-to-light ratios of 0.07$\pm$0.02, 0.09$\pm$0.02, and 0.20$\pm$0.07\,M$_\mathrm{\odot}$/L$_\mathrm{\odot}$ were  derived for stellar populations characterised by 10, 20, and 50$-$100\,Myr constant SFRs, respectively. For an ageing population represented by instantaneous bursts from 200 to 400\,Myr old, the M/L  is 1.30$\pm$0.47\,M$_\mathrm{\odot}$/L$_\mathrm{\odot}$. 

The stellar clumps are very massive, with masses ranging from 3 to 5 $\times$ 10$^9$ M$_{\odot}$ for stellar populations with ages 50-100 Myr. These masses could be smaller by a factor of two to three for younger populations and a factor of 6.5 higher for populations that are a few to several hundred million years old (see Table~\ref{tab:Mass_F1000W_reg} for stellar masses for each F1000W clump and different stellar populations). Clumps dominated by these old populations are unrealistic, as their predicted stellar mass would be well above their measured dynamical mass (see Table~\ref{tab:Mass_CII_reg}). We considered the masses derived for a 50-100 Myr as the baseline for the detailed comparison with the dynamical and gas mass in the clumps.

Based on the structural analysis presented in section~\ref{sec:4.2.}, the stellar mass for the [\ion{C}{II}] clumps was estimated as the sum of the mass of the F1000W clumps that are considered as physically associated.  This correspondence between clumps, indicated in Table~\ref{tab:Mass_CII_reg}, was only applied to those [\ion{C}{II}] regions that presented counterpart emission in the F1000W image (i.e. E2, E3, W4, W5, and W6). 
Each of the [\ion{C}{II}] clumps have very high stellar masses with 3 to 9 $\times$ 10$^9$ M$_\odot$. However, the stellar mass appears to be a low fraction (28-45$\%$) of the dynamical mass, except for clump E3 in the E galaxy. The mass within the clumps appears to be dominated by the gas component, especially in the clumps in the W galaxy (see Table~\ref{tab:Mass_CII_reg}). Only when considering stellar populations dominated by aged stars (200$-$400\,Myr old) could the dynamical mass be dominated by the stars in most of the clumps. However, the stellar masses under this last  assumption would increase the mass up to unrealistic values of 2 to 6 $\times$10$^{10}$\,M$_\mathrm{\odot}$ per clump.

\begin{figure}
   \includegraphics[width=\linewidth]{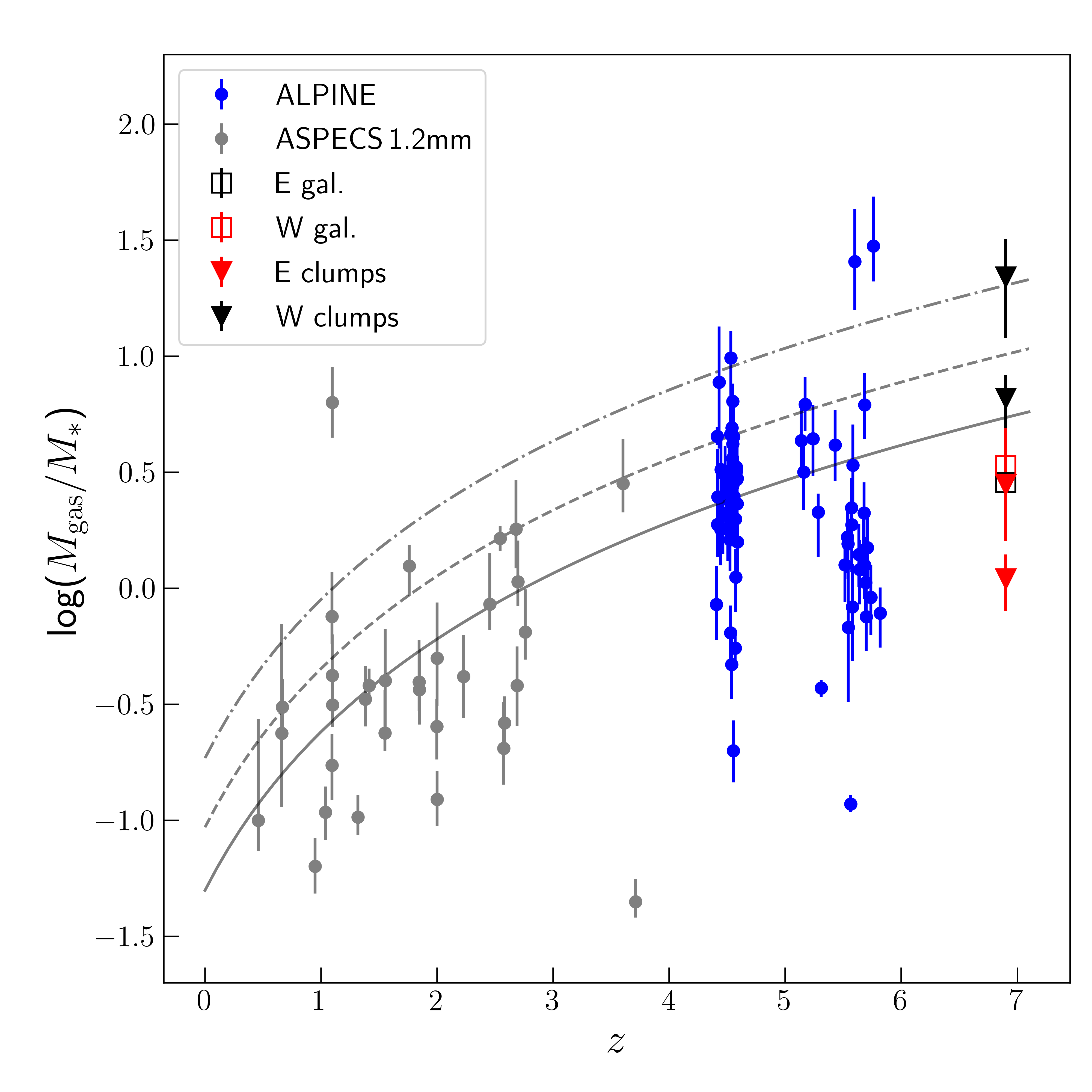}
      \caption{Gas fraction vs redshift. This figure shows the evolution of the gas mass fraction (i.e. $M_\mathrm{gas}/M_\mathrm{*}$) with the redshift. Black and red triangles represent the values derived for the E and W SPT0311-058 clumps, respectively, presented in Table~\ref{tab:Mass_CII_reg}. Red and black empty squares display the $M_\mathrm{gas}/M_\mathrm{*}$ ratio obtained using the gas mass derived by \citet{Marrone+18} and the stellar masses computed in this work via SED fitting (see Sec.~\ref{sec:4.1.}) for the E and W galaxies, respectively. For comparison, galaxies at intermediate redshifts ($z\sim$1$-$3) from the ASPECS 1.2mm sample presented in \citep{Aravena+20} are included. The higher redshift range ($z\sim$4.4$-$6) is covered with the ALPINE sample, for which we calculate the gas mass following the $L_\mathrm{CII}-M_\mathrm{gas}$ relation introduced by \citet{Zanella+18}. The black continuum, dashed, and dashed–dotted lines show the gas fraction evolution following \citet{Liu+19} for a galaxy with a stellar mass of $M_*$=5$\times$10$^{10}$\,M$_\odot$ and an SFR that is one, three, and ten times the one expected for the SFR main sequence, respectively.}
     
     \label{fig:frac_gas}
\end{figure}

Assuming the dynamical mass in the clumps is dominated by the gas component, the ratio of the gas (traced by the \CII{} line) to stellar mass in the E and W clumps appears to have significant differences (see Table~\ref{tab:Mass_CII_reg}). For the E galaxy, this ratio is between one and three. The clumps in the W galaxy are more extreme, with gas to stellar mass ratios of seven to 22 (but we note that gas masses derived for W clumps are higher by a factor of three to seven than the dynamical mass, see Table~\ref{tab:Mass_CII_reg}). These ratios (see Figure~\ref{fig:frac_gas}) are much higher, in particular for the W clumps, than those measured ($M_\mathrm{gas}$/$M_*${}$\leq$1) in star-forming galaxies at redshifts 1 to 3 \citep{Aravena+20}, but they are in the upper range of the ratios measured at higher redshifts ($z\sim$ 4.4 to 5.9) in [\ion{C}{II}]-detected galaxies \citep{Dessauges-Zavadsky+20}. However, the mass ratios measured in the SPT0311-58 clumps have to be treated with caution  due to potential uncertainties related to the dependence of the ratio on the assumed ages for the stellar population and the lower limits measured for the internal (nebular and stellar) extinction. Clumps with stellar populations older than 50-100 Myr are more massive, and therefore the ratio of the gas to stars would be lower than the values above. Even if no old stellar populations were present in the clumps, the derived Pa$\alpha$/H$\alpha$ nebular and corresponding stellar extinctions are lower limits of the true internal extinction. Extinction corrections applied to the 10$\mu$m flux (i.e. near-infrared rest-frame) would increase the amount of obscured stellar mass for the same stellar population. Some evidence that this could be happening, at least in the W galaxy, comes from the $M_\mathrm{gas}/M_*$ ratio derived using the gas mass from the 160$\mu$m emission ($M_\mathrm{gas}$= 4 and 27 $\times$ 10$^{10}$\,M$_{\odot}$ for the E and W galaxies, \citealt{Marrone+18}) and the stellar mass from the SED fitting carried out in this work (see values in Table \ref{tab:prop_SED}). For this scenario, the mass ratio becomes 2.9 (E) and 3.4 (W), respectively (see Figure~\ref{fig:frac_gas}). While the value in the E galaxy agrees with those derived for the clumps, it is lower by a factor of two to seven in the W galaxy. The 160$\mu$m- and SED-based mass ratios are more moderate and within the range measured in $z\sim$ 4.4$-$5.9 galaxies (Figure~\ref{fig:frac_gas}). Predictions of the evolution of gas mass fractions as a function of redshift \citep{Liu+19} expect $M_\mathrm{gas}/M_\mathrm{*}$ ratios of ten to 21 for massive starbursts (i.e. galaxies with SFRs three to ten times above the so-called main sequence of star-forming galaxies) at redshift 7. Within the uncertainties and caveats considered above, the mass ratios derived in both the clumps and the entire galaxies of the SPT0311-58 system are well below these limits and within the range of the values measured at lower redshifts. This would imply a flattening in the evolution of the gas mass fraction in massive star-forming galaxies from redshifts of about 4 up to redshifts of 7, even for luminous and massive starbursts such as SPT0311-58.

\begin{table}
\caption{Stellar masses for MIRI/F1000W clumps with different star-forming histories.}             
\label{tab:Mass_F1000W_reg}      
\centering          
\begin{tabular}{c c c c}  
\hline\hline       
Region & M$_\mathrm{*}$[10\,Myr] & M$_\mathrm{*}$[20\,Myr]  & M$_\mathrm{*}$[50$-$100\,Myr] \\ 
 & (10$^{9}$M$_\mathrm{\odot}$) &  (10$^{9}$M$_\mathrm{\odot}$) & (10$^{9}$M$_\mathrm{\odot}$)  \\
 (1) & (2) & (3) & (4) \\
\hline                    
EA &  1.4$\pm$0.5 &  1.9$\pm$0.7 &  4.2$\pm$1.8 \\
EB &  1.5$\pm$0.4 &  2.0$\pm$0.6 &  4.4$\pm$1.7 \\
EC &  1.6$\pm$0.5 &  2.2$\pm$0.7 &  4.9$\pm$1.9 \\
ED &  1.1$\pm$0.3 &  1.5$\pm$0.5 &  3.2$\pm$1.3 \\
EE &  1.1$\pm$0.3 &  1.5$\pm$0.5 &  3.2$\pm$1.3 \\
WA &  1.1$\pm$0.4 &  1.4$\pm$0.6 &  3.2$\pm$1.5 \\
WB &  1.0$\pm$0.3 &  1.4$\pm$0.4 &  3.1$\pm$1.2 \\
WC &  1.3$\pm$0.4 &  1.8$\pm$0.5 &  4.0$\pm$1.5 \\
WD &  1.5$\pm$0.4 &  2.0$\pm$0.6 &  4.5$\pm$1.6 \\
\hline                  
\end{tabular}
\tablefoot{Columns (2), (3), and (4) give the extinction-corrected stellar mass based on the $L_\mathrm{J}$ luminosity and constant M/L ratios for a constant SFR over 10\,Myr, 20\,Myr, and 50$-$100\,Myr, respectively. Extinction has been corrected assuming an $A_\mathrm{V}$ of 2.4 and 2.2 for the E and W stellar components (see $\S$\ref{sec:4.3.}) and Calzetti's extinction law \citep{Calzetti+00}.}
\end{table}

\begin{table*}
\caption{Properties of SPT0311-58 clumps identified in the ALMA rest-frame 160$\mu$m continuum image.}           
\label{tab:Mass_CII_reg}      
\centering          
\begin{tabular}{c c c c c c c c c}  
\hline\hline       
Region & F1000W clump & $M_\mathrm{dyn}$ & $M_\mathrm{gas}$ & SFR$_\mathrm{[CII]}$ & $M_\mathrm{*}$[50$-$100\,Myr] & $M_\mathrm{dyn}$/$M_\mathrm{*}$ & $M_\mathrm{gas}$/$M_\mathrm{*}$ & sSFR  \\ 
 & & (10$^{9}$M$_\mathrm{\odot}$) &  (10$^{9}$M$_\mathrm{\odot}$)  & (M$_\mathrm{\odot}$yr$^{-1}$) & (10$^{9}$M$_\mathrm{\odot}$) &  &   & (Gyr$^{-1}$) \\
 (1) & (2) & (3) & (4) & (5) & (6) & (7) & (8) & (9) \\ 
\hline                    
E2    &          EA &   14\oddpm{14}{7} &   12 &    250\oddpm{120}{80} &   4.2$\pm$1.8 &  3.5\oddpm{3.8}{2.3} &    2.8$\pm$1.2 &   60\oddpm{38}{32} \\
E3    &       EB+EC &     9\oddpm{9}{5} &   10 &    230\oddpm{110}{80} &   9.3$\pm$2.5 &  1.0\oddpm{1.0}{0.6} &    1.1$\pm$0.3 &   25\oddpm{14}{11} \\
W4    &          WA &     9\oddpm{8}{4} &   68 &     140\oddpm{70}{50} &   3.2$\pm$1.5 &  2.7\oddpm{3.0}{1.8} &  22$\pm$10 &   44\oddpm{30}{26} \\
W5+W6 &       WC+WD &   18\oddpm{13}{7} &   56 &    300\oddpm{100}{70} &   8.5$\pm$2.2 &  2.2\oddpm{1.7}{1.0} &    6.6$\pm$1.7 &   35\oddpm{15}{12} \\
E$^*$     &      galaxy &  40\oddpm{21}{10} &   28 &   770\oddpm{190}{130} &  19.9$\pm$3.6 &  2.0\oddpm{1.1}{0.6} &    1.4$\pm$0.3 &   39\oddpm{12}{10} \\
W$^*$   &      galaxy &  72\oddpm{28}{14} &  227 &  1760\oddpm{330}{220} &  14.8$\pm$2.9 &  4.9\oddpm{2.1}{1.3} &   15$\pm$3 &  119\oddpm{32}{28} \\
\hline                  
\end{tabular}
\tablefoot{Column 1 displays the \CII{} regions considered to compute the parameters. E$^*$ and W$^*$ represent the sum of all the [\ion{C}{II}] clumps for each galaxy. Column 2 shows the clump correspondence between the regions identified in the ALMA rest-frame 160$\mu$m and MIRIM F1000W images. The last two rows represent the total values for each galaxy, derived by summing all the clumps. Columns 3, 4, and 5 give the intrinsic dynamical, gas masses, and SFR derived from the [\ion{C}{II}]158$\mu$m ALMA spectral imaging \citep{Spilker+22}. Column 6 gives the extinction-corrected stellar masses derived from the MIRI F1000W imaging for a 50$-$100\,Myr constant star formation presented in Table~\ref{tab:Mass_F1000W_reg}. Considering stellar populations created by a constant star formation of $\sim$10\,Myr or instantaneous burst of 200$-$400\,Myr would yield M$_\mathrm{*}$ different by a factor of $\sim$0.33 and $\sim$6.5, respectively. Columns 7 and 8 show the ratio between the stellar mass computed in this work and the dynamical and gas masses derived from the [\ion{C}{II}]. Column 9 displays the  sSFR (SFR/$M_*$) derived using the SFR values computed by \citet{Spilker+22}.}
\end{table*}

\subsection{The starburst clumpy nature of SPT0311-58 and the formation of massive galaxies in the early Universe}
\label{sec:4.7.}

\begin{figure*}
   \includegraphics[width=\linewidth]{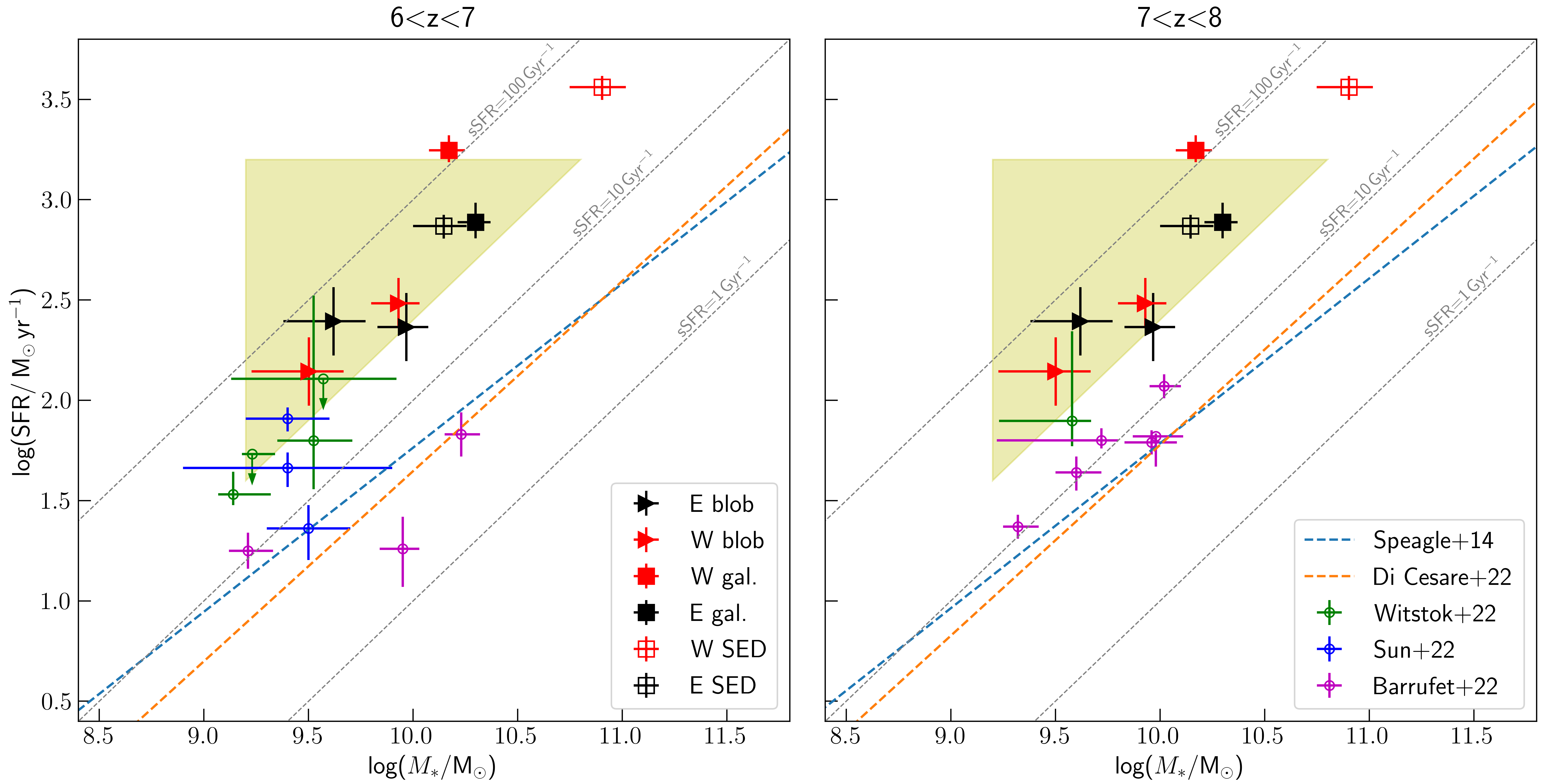}
      \caption{Star formation rate vs stellar mass. This figure shows the position of the F1000W clumps associated with the [\ion{C}{II}] clumps in the SFR versus stellar mass plane. The stellar masses are extinction-corrected and assume a continuum star formation of $\sim$50$-$100\,Myr. The SFR is from the SFR$_\mathrm{CII}$ derived by \citet{Spilker+22}. The values for the E and W galaxy clumps are displayed as black and red triangles, respectively. Filled squares represent the total values for the E and W galaxies using the F1000W masses and [\ion{C}{II}] SFRs. Empty squares are the analogous points using the M$_*$ and SFR derived from the SED fitting (see Table~\ref{tab:prop_SED}). Grey dashed lines represent constant sSFR.  blue dashed and orange dashed lines in both panels represent the SFMS derived by \citet{Speagle+14} and \citet{diCesare+22}, respectively, at $z${}$\sim$6.5 and $z${}$\sim$7.5. For a direct comparison, we transformed the MS by \citet{diCesare+22} from a Salpeter to Kroupa IMF, multiplying the SFRs and M$_\mathrm{*}$ by 0.67 and 0.66, respectively \citep{Madau&Dickinson+14}. Yellow-shaded area represents the `starburst region' at $z${}$\sim$4.3 defined by \cite{Caputi+17}. Green, blue, and magenta stars represent the results by \citet{Witstok+22}, \citet{Sun+22}, and \citet{Barrufet+22}, respectively.}
     
     \label{fig:SFMS}
\end{figure*}

The F1000W clumps identified in the SPT0311-58 have stellar masses similar to that of galaxies recently detected at redshifts 6 to 8 with {\it JWST} and ALMA (e.g \citealt{Witstok+22}, \citealt{Sun+22} and \citealt{Barrufet+22}). To establish a comparison with these high-$z$ galaxies, the position of the F1000W clumps associated with the [\ion{C}{II}] clumps in the $M_\mathrm{*}-$SFR plane (Figure~\ref{fig:SFMS}) was derived taking the [\ion{C}{II}]158$\mu$m SFR measurements (see Table~\ref{tab:Mass_CII_reg}). For a 50$-$100\,Myr constant star formation, the clumps were characterised by an sSFR (SFR/$M_{*}$) of 25 to 60\,Gyr$^{-1}$. This sSFR is a factor $\sim$3$-$10 times higher than expected for main-sequence star-forming galaxies of similar mass at $z$=6$-$8, as predicted by hydrodynamical cosmological simulations of dusty galaxies \citep{diCesare+22} or extrapolations from lower redshifts \citep{Speagle+14}. In addition, the clumps in SPT0311-58 appear to be in a starbursting status when compared with other equally massive star-forming galaxies at similar redshifts recently detected by {\it JWST} (Figure~\ref{fig:SFMS}, \citealt{Sun+22} and \citealt{Barrufet+22}). Each of the F1000W clumps, as well as at least the E galaxy, occupy a region in the $M_\mathrm{*}-$SFR plane similar to that of massive (1.6-60 $\times$ 10$^9$ M$_{\odot}$) starbursts at $z$\,$\sim$4$-$5 characterised by sSFR of 40 to 100 Gyr$^{-1}$ \citep{Caputi+17,Caputi+21}. Therefore, SPT0311-58 is an extreme system formed by two starburst galaxies each consisting of several massive starbursting clumps in a region of few kiloparsecs in size. 

The presence of massive kpc-size star-forming clumps has been predicted in cosmological simulations of gas dominated discs that are fed by cosmological cold streams \citep{Dekel+09,Ceverino+10,Mandelker+14}. The clumps are produced as a natural consequence of instabilities in the gas dominated systems that have a quasi-continuous supply of raw material from the cold streams of gas. Under the scenario of disc instabilities, giant clumps that have masses of a few percent of the mass of the disc are predicted to form \citep{Ceverino+10}. SPT0311-58 is a very gas-rich system with intrinsic gas masses of 3.1$\pm$2.7$\times$10$^{11}$\,M$_\mathrm{\odot}$ and 5.4$\pm$3.4$\times$10$^{11}$\,M$_\mathrm{\odot}$ in the E and W galaxies \citep{Jarugula+21}. Therefore, clumps with typical masses of about a few 10$^9$ to 10$^{10}$\,M$_\mathrm{\odot}$ are expected. These masses are in agreement with the stellar and dynamical masses derived for the clumps. Following these cosmological simulations, the massive clumps will migrate inwards in about two orbital periods, forming a spheroidal galaxy \citep{Dekel+09}. Since the clumps are distributed in small volumes of a few kiloparsecs in radius and the orbital velocities correspond to about 200$-$300\,km\,s$^{-1}$, the timescale is short (i.e. <100\,Myr), and a bulge grows in a few to several hundred million years. In addition, the feeding of a black hole and its transformation into a bright AGN has also been considered as part of the evolutionary scenario of clumpy discs in high redshift galaxies \citep{Bournaud+11,Bournaud+12}. Following this overall scenario, the SPT0311-58 system could be in an evolutionary phase that would rapidly evolve into a massive spheroid, potentially feeding a central black hole and therefore generating an infrared-luminous quasi-stellar object similar to those already detected at redshifts of 6 and above \citep{Venemans+20,Walter+22}.

Massive clumps could also be formed in major mergers, as indicated by the observed increasing clumpiness of mergers that are identified as starbursts in intermediate redshift galaxies \citep{Calabro+18}. This scenario is further supported by the hydrodynamical simulations of mergers of gas discs that confirm the formation of star-forming clumps during the evolution of the merger as opposed to isolated discs \citep{Calabro+19}. Merger scenarios \citep{Bournaud+11} predict SFRs with peaks of 1000-2000 M$_{\odot}$ yr$^{-1}$ and clumps that become more massive and more dense during the evolution of the merger. SPT0311-58 is a system formed by two galaxies separated by $\sim$7\,kpc and that are likely interacting; thus, a merger scenario could be at play here.

In summary, SPT0311-58 is a system composed of two massive gas-rich interacting galaxies with a clumpy morphology consisting of several massive gas and stellar clumps within a radius of a few kpc. Independent of the dominant scenario, be it disc instabilities, mergers, or a combination of both, new simulations with high angular resolution (at least 100 pc) exploring the formation and evolution of such  galaxies as SPT0311-58 in the early Universe (i.e. at ages of less than 700-800\,Myr) are required to understand their evolution considering the physical characteristics measured by JWST and ALMA. Also, additional ALMA and JWST observations are needed to put better constraints on the internal extinction, ages of the stellar population, and presence of an obscured AGN as well as to obtain  more accurate stellar and gas masses in the clumps. Some of these additional observations will be taken soon with JWST. 

\section{Summary}
\label{sec:5.}

SPT0311-58 consists of two gas-rich galaxies at redshift 6.9 separated by about 8 kpc and experiencing a strong starburst.
This paper has presented mid-infrared sub-arcsec imaging and spectroscopy of the SPT0311-58 system taken with MIRI/JWST. The data provide the first resolved kpc-scale rest-frame near-infrared light of an extreme dusty, infrared-luminous star-forming system within the Epoch of Reionization. The data enable the exploration of the stellar and ionised gas structure of the galaxies that form the system. The study also combines the new MIRI data with archival high angular resolution ALMA imaging to derive additional physical properties, such as the SFR, the ratio of the stellar mass to dynamical and gas masses, and the specific SFRs. The main results are summarised below: 

\begin{itemize} 
    
    \item The stellar light of the two galaxies (E and W) that belong to the system shows a clumpy structure with a total of nine clumps (five in the E and four in W, plus one tentative clump). The clumps appear as unresolved sources and within regions of a few kpc in size.

    \item The clumps are very massive, with stellar masses in the range of 3 to 5 $\times$10$^{9}$\,M$_\mathrm{\odot,}$ assuming a constant star formation of 50 to 100\,Myr.  The stellar mass in the massive clumps represents a fraction, 20\% up to 100\% but with an average of 43\%, of their dynamical mass, implying that the mass of the clumps is dominated by gas. 
    
    \item The gas to stellar mass ratios derived in the clumps and entire galaxies of the system appear to be consistent with values (1-3) of [\ion{C}{II}]-detected galaxies at redshifts above 4. Which is in agreement with a flattening in the evolution of the gas mass fraction, even in massive starbursts at redshift 7 such as SPT0311-58. Extreme ratios measured in some W clumps ($M_\mathrm{gas}$/$M_*$\,$\sim$\,7 to 22) could be due to a combination of effects, including internal extinction, the presence of stellar populations older than 100\,Myr, and uncertainties in the gas estimates.

    \item The stellar and interstellar medium morphology of the two galaxies shows a complex structure. Some of the stellar clumps detected in the MIRI images are spatially offset from those already identified in the rest-frame 160$\mu$m continuum and [\ion{C}{II}]158$\mu$m line. In addition, some far-infrared clumps (e.g. W1, W2, W8, and E1) are not detected in near-infrared light. Other stellar clumps detected with MIRI (e.g. ED and EE) do show UV emission (HST) but not far-infrared emission.  This complex morphology is likely due to the combination of an intrinsic clumpy stellar structure with several independent clumps forming stars at different rates (i.e. different sSFR) and a patchy dust distribution producing internal differential extinction effects, very similar to what is observed in low-$z$ ultra-luminous infrared galaxies at (sub)kpc scales.  
    
    \item The ionised interstellar medium is traced by the Pa$\alpha$ emission line present in the two galaxies of the system, with no detection of H$\alpha$. This implies a large internal nebular extinction in the two galaxies with lower limits of 4.2 mag and 3.9 mag in the visual ($A_\mathrm{V}$) for the E and W galaxies, respectively. These large extinctions agree with values measured in low- and intermediate-$z$ ultra-luminous galaxies. Nebular extinction based on SED fitting provides larger values, $A_\mathrm{V}$\,=\,5.6 (E) and 10.0 (W). The SED-based extinction towards stars older than 10 Myr is lower and amounts to 1.3 and 5 visual magnitudes.
    
    \item The lower limits of the ongoing SFR traced by the extinction corrected Pa$\alpha$ line corresponds to 383 and 230\,M$_\mathrm{\odot}$\,yr$^{-1}$ for the E and W galaxies. This represents 50\% and 13\% of the SFR derived from the [\ion{C}{II}]158$\mu$m line. These amounts are similar to those measured in typical low-$z$ LIRGs (E) and in extreme dust-enshrouded ULIRGs (W).
    
     \item The sSFR in the stellar clumps ranges from 25 to 60\,Gyr$^{-1}$ and is three to ten times larger than the predicted values for main-sequence star-forming galaxies of similar mass at redshifts 6 to 8 in cosmological simulations of dusty galaxies \citep{diCesare+22} and  recently detected by JWST. Thus, SPT0311-58 is a starburst system even when compared with typical massive star-forming galaxies at such high redshifts. 

\end{itemize}

The stellar mass and spatial distribution of several massive star-forming clumps within a few kpc in each of the galaxies that form the SPT0311-58 system is consistent with the scenario of the formation of massive clumps in high-$z$ gas-rich discs.  The dynamical evolution of the system should be very fast, with the clumps moving inwards in less than 100\,Myr and producing a massive spheroid \citep{Dekel+09}. In addition, a central BH could be activated and produce a luminous AGN very similar to most of the quasi-stellar objects at redshifts above 6 where a luminous AGN coexists with a powerful starburst in the nucleus. Thus, SPT0311-58 could be exhibiting the evolutionary phase immediately previous to the infrared phase of a luminous quasi-stellar object. Alternative scenarios predicting the formation and evolution of massive clumps as a consequence of  interaction and merger can also be relevant in this galaxy. The combination of high angular resolution JWST and ALMA data and hydrodynamical cosmological simulations of massive structures in the early Universe are required to further illuminate the formation and subsequent evolution of systems like SPT0311-58 during the first Gyr after the Big Bang.

\begin{acknowledgements}
The  authors  thank  to the  anonymous  referee  for  useful  comments. J.A-M., A.C-G., L.C., A.L. acknowledge support by grant PIB2021-127718NB-100, P.G.P-G. by grant PGC2018-093499-B-I00. A.A-H. by grant PID2021-124665NB-I00 from the Spanish Ministry of Science and Innovation/State Agency of Research MCIN/AEI/10.13039/501100011033 and by “ERDF A way of making Europe”. F.W. and M.N. acknowledge support from the ERC Advanced Grant 740246 (Cosmic\_Gas), A.B. and G.O. acknowledges support from the Swedish National Space Administration (SNSA).
O.I. acknowledges the funding of the French Agence Nationale de la Recherche for the project iMAGE (grant ANR-22-CE31-0007),
J.H. was supported by a VILLUM FONDEN Investigator grant (project number 16599). K.I.C. acknowledges funding from the Netherlands Research School for Astronomy (NOVA) and the Dutch Research Council (NWO) through the award of the Vici Grant VI.C.212.036. J.P.P. acknowledges financial support from the UK Science and Technology Facilities Council, and the UK Space Agency. T.P.R. would like to acknowledge support from the ERC under advanced grant 743029 (EASY). The Cosmic Dawn Center (DAWN) is funded by the Danish National Research Foundation under grant No. 140

      
The work presented is the effort of the entire MIRI team and the enthusiasm within the MIRI partnership is a significant factor in its success. MIRI draws on the scientific and technical expertise of the following organisations: Ames Research Center, USA; Airbus Defence and Space, UK; CEA-Irfu, Saclay, France; Centre Spatial de Liége, Belgium; Consejo Superior de Investigaciones Científicas, Spain; Carl Zeiss Optronics, Germany; Chalmers University of Technology, Sweden; Danish Space Research Institute, Denmark; Dublin Institute for Advanced Studies, Ireland; European Space Agency, Netherlands; ETCA, Belgium; ETH Zurich, Switzerland; Goddard Space Flight Center, USA; Institute d'Astrophysique Spatiale, France; Instituto Nacional de Técnica Aeroespacial, Spain; Institute for Astronomy, Edinburgh, UK; Jet Propulsion Laboratory, USA; Laboratoire d'Astrophysique de Marseille (LAM), France; Leiden University, Netherlands; Lockheed Advanced Technology Center (USA); NOVA Opt-IR group at Dwingeloo, Netherlands; Northrop Grumman, USA; Max-Planck Institut für Astronomie (MPIA), Heidelberg, Germany; Laboratoire d’Etudes Spatiales et d'Instrumentation en Astrophysique (LESIA), France; Paul Scherrer Institut, Switzerland; Raytheon Vision Systems, USA; RUAG Aerospace, Switzerland; Rutherford Appleton Laboratory (RAL Space), UK; Space Telescope Science Institute, USA; Toegepast- Natuurwetenschappelijk Onderzoek (TNO-TPD), Netherlands; UK Astronomy Technology Centre, UK; University College London, UK; University of Amsterdam, Netherlands; University of Arizona, USA; University of Cardiff, UK; University of Cologne, Germany; University of Ghent; University of Groningen, Netherlands; University of Leicester, UK; University of Leuven, Belgium; University of Stockholm, Sweden; Utah State University, USA. A portion of this work was carried out at the Jet Propulsion Laboratory, California Institute of Technology, under a contract with the National Aeronautics and Space Administration. We would like to thank the following National and International Funding Agencies for their support of the MIRI development: NASA; ESA; Belgian Science Policy Office; Centre Nationale D'Etudes Spatiales (CNES); Danish National Space Centre; Deutsches Zentrum fur Luft-und Raumfahrt (DLR); Enterprise Ireland; Ministerio De Econom\'ia y Competitividad; Netherlands Research School for Astronomy (NOVA); Netherlands Organisation for Scientific Research (NWO); Science and Technology Facilities Council; Swiss Space Office; Swedish National Space Board; UK Space Agency. 

This work is based on observations made with the NASA/ESA/CSA James Webb Space Telescope. The data were obtained from the Mikulski Archive for Space Telescopes at the Space Telescope Science Institute, which is operated by the Association of Universities for Research in Astronomy, Inc., under NASA contract NAS 5-03127 for \textit{JWST}; and from the \href{https://jwst.esac.esa.int/archive/}{European \textit{JWST} archive (e\textit{JWST})} operated by the ESDC.

This paper makes use of the following ALMA data: ADS/JAO.ALMA\#2016.1.01293.S and ADS/JAO.ALMA\#2017.1.01493.S. ALMA is a partnership of ESO (representing its member states), NSF (USA) and NINS (Japan), together with NRC (Canada), MOST and ASIAA (Taiwan), and KASI (Republic of Korea), in cooperation with the Republic of Chile. The Joint ALMA Observatory is operated by ESO, AUI/NRAO and NAOJ. The National Radio Astronomy Observatory is a facility of the National Science Foundation operated under cooperative agreement by Associated Universities, Inc. 

This research made use of Photutils, an Astropy package for detection and photometry of astronomical sources \citep{larry_bradley_2022_6825092}.

\end{acknowledgements}

\bibliographystyle{aa} 
\bibliography{bibliography.bib} 

\end{document}